\documentclass{emulateapj}

\usepackage{array} 
\usepackage{hyperref}
\usepackage{graphicx}
\usepackage{color}
\usepackage{multirow} 
\usepackage{url}
\bibpunct{(}{)}{;}{a}{}{,}

\shorttitle{The Yale-Potsdam Stellar Isochrones (YaPSI)}
\shortauthors{Spada et al.}

\begin{document}

\title{The Yale-Potsdam Stellar Isochrones (YaPSI)}

\author{F. Spada}
\affiliation{Leibniz-Institut f\"ur Astrophysik Potsdam (AIP), An der Sternwarte 16, D-14482, Potsdam, Germany}

\author{P. Demarque}
\affiliation{Department of Astronomy, Yale University, New Haven, CT 06520-8101, USA}

\author{Y.--C. Kim}
\affiliation{Yonsei University Observatory and Astronomy Department, Yonsei University, Seoul 120-749, South Korea}

\author{T. S. Boyajian}
\affiliation{Physics and Astronomy, Louisiana State University, Baton Rouge, LA 70803, USA}
\affiliation{Department of Astronomy, Yale University, New Haven, CT 06520-8101, USA}

\author{J. M. Brewer}
\affiliation{Department of Astronomy, Yale University, New Haven, CT 06520-8101, USA}

\email{fspada@aip.de}

\begin{abstract}
We introduce the Yale-Potsdam Stellar Isochrones (YaPSI), a new grid of stellar evolution tracks and isochrones of solar-scaled composition.
In an effort to improve the Yonsei-Yale database, special emphasis is placed on the construction of accurate low-mass models ($M_* < 0.6 \, M_\odot$), and in particular of their mass-luminosity and mass-radius relations, both crucial in characterizing exoplanet-host stars and, in turn, their planetary systems. 
The YaPSI models cover the mass range $0.15$ to $5.0\, M_\odot$, densely enough to permit detailed interpolation in mass, and the metallicity and helium abundance ranges [Fe/H] $=-1.5$ to $+0.3$, and $Y_0=0.25$ to $0.37$, specified independently of each other (i.e., no fixed $\Delta Y/ \Delta Z$ relation is assumed). 
The evolutionary tracks are calculated from the pre-main sequence up to the tip of the red giant branch. 
The isochrones, with ages between $1$ Myr and $20$ Gyr, provide UBVRI colors in the Johnson-Cousins system, and JHK colors in the homogeneized Bessell \& Brett system, derived from two different semi-empirical $T_{\rm eff}$-color calibrations from the literature. 
We also provide utility codes, such as an isochrone interpolator in age, metallicity, and helium content, and an interface of the tracks with an open-source Monte Carlo Markov-Chain tool for the analysis of individual stars.  
Finally, we present comparisons of the YaPSI models with the best empirical mass-luminosity and mass-radius relations available to date, as well as isochrone fitting of well-studied stellar clusters.
\end{abstract}

\keywords{}

\def\aj{{AJ}}                   
\def\araa{{ARA\&A}}             
\def\apj{{ApJ}}                 
\def\apjl{{ApJ}}                
\def\apjs{{ApJS}}               
\def\aap{{A\&A}}                
\def\apss{{Ap\&SS}}          
\def\aapr{{A\&A~Rev.}}          
\def\aaps{{A\&AS}}              
\def\mnras{{MNRAS}}             
\def\nat{{Nature}}              
\def\ssr{{Space~Sci.~Rev.}}

\section{Introduction}
\label{introduction}

The purpose of this paper is to present the Yale-Potsdam Stellar Isochrones (YaPSI)\footnote{Yale Astro web page: \url{http://www.astro.yale.edu/yapsi/}; AIP web page: \url{http://vo.aip.de/yapsi/}}, a new grid of stellar evolutionary calculations covering the low and intermediate mass regimes from the early pre-main sequence to the end of the red giant branch phase. 
Our grid features a dense coverage of the mass range $0.15$ to $5.00\, M_\odot$ that permits accurate interpolation in mass. 
A broad range of solar-scaled initial compositions is available, with metallicity in the range [Fe/H]$_0=-1.5$ to $+0.3$, and helium content $Y_0 = 0.25$ to $0.37$.

As a follow-up and extension of a previous paper \citep{Spada_ea:2013}, a special emphasis of this work is on the construction of accurate mass-luminosity and mass-radius relations, in particular for low-mass stars ($M_* \lesssim 0.6 \, M_\odot$).
The focus of the YaPSI models tries to reflect the novel or recently renewed interests of the stellar physics community, for example in the characterization of open clusters (e.g., the WIYN Open Cluster Study, \citealt{Mathieu:2000}, or the Monitor Project, \citealt{Aigrain_ea:2007}), and even of field stars (e.g., gyrochronology: \citealt{Barnes:2003, Barnes_ea:2016}; asteroseismology: \citealt{Basu_ea:2010, Chaplin_Miglio:2013}; direct interferometric measurement of stellar radii: \citealt{Boyajian_ea:2012}), in the wake of space missions such as CoRoT and {\it Kepler}, and at the dawn of the Gaia era \citep[Gaia Data Release 1 has just been made public:][]{Lindegren_ea:2016}.

The theory of low-mass stars and substellar objects has a long history and presents peculiar challenges, most notably the treatment of the atmospheric boundary conditions and the equation of state \citep[see, for example,][]{Limber:1958,Burrows_ea:1989,Chabrier_Baraffe:1997, Allard_ea:1997,Chabrier_Baraffe:2000}.
Decisive improvement in these pieces of input physics were included in the Lyon models of \citet{Baraffe_ea:1998}, which have been for a long time one of the main references for theoretical models of low and very-low-mass stars.
Since then, these authors have updated their calculations \citep{Baraffe_ea:2015}, while other groups have expanded and improved their grids towards the low-mass regime (Dartmouth: \citealt{Dotter_ea:2008}; Padova/PARSEC: \citealt{Chen_ea:2014}; Victoria-Regina: \citealt{Vandenberg_ea:2014}; MESA: \citealt{Choi_ea:2016}, among others).

In this sense, the YaPSI release is intended as an update and an extension of the Yonsei-Yale isochrones (YY hereafter: \citealt{Yi_ea:2001, Kim_ea:2002, Yi_ea:2003,Demarque_ea:2004}), featuring models of low-mass stars that incorporate the latest improvements in the relevant input physics, and homogenizing the results of an early release \citep{Spada_ea:2013} with the YY database\footnote{\url{http://www.astro.yale.edu/demarque/yyiso.html}}.   
Care has been taken, whenever possible, to preserve compatibility with the utility codes developed for YY.

Some of the applications for which we envisage that the YaPSI models are especially well-suited are highlighted as follows.

The mass-luminosity relation (MLR in the following) and the mass-radius relation play a fundamental role in stellar astrophysics \citep[see, e.g.,][]{Schwarzschild:1958}. 
They are especially important for exoplanet-host stars: the planetary radii are determined relative to (and therefore only as accurately as) the radius of their host star \citep[e.g.,][]{Charbonneau_ea:2000}.  

The MLR provides both a stringent test of the input physics included in stellar models \citep{Andersen:1991}, and a crucial link between the luminosity of a star and its mass \citep{Henry:2004}.
The best empirical MLR for low-mass stars available so far, the result of more than 20 years of observational work, has just been published \citep{Benedict_ea:2016}.
The YaPSI models are compared with this exquisitely precise empirical relation in Section \ref{ssec_benedict}.

The mass-radius relation has received a great deal of attention in the context of the so-called ``radius inflation problem". 
When sufficiently accurate measurements of stellar masses and radii are simultaneously available, stars of mass $\lesssim 0.7 \, M_\odot$ are found to have radii inflated by about $3$--$5\%$ with respect to model predictions \citep[e.g.,][and references therein; see also Section \ref{ssec_rinfl}]{Torres_ea:2010,Feiden_Chaboyer:2012a,Spada_ea:2013}.
The explanation of this discrepancy has been intriguingly linked by many authors to some piece of missing physics in the models, such as magnetic activity \citep{Morales_ea:2010}, magnetic fields \citep{Feiden_Chaboyer:2012b,MacDonald_Mullan:2013}, or, indirectly, to rotation \citep{Somers_Pinsonneault:2015, Lanzafame_ea:2016}. 

The YaPSI tracks also include rotation-related stellar parameters, such as the moments of inertia or the convective turnover time scale, that are useful in studies of the angular momentum evolution in stellar interiors \citep[see, e.g.,][]{Spada_ea:2011, Penev_ea:2012, Gallet_Bouvier:2013, Lanzafame_Spada:2015}, and in applying the stellar age-dating technique of gyrochronology \citep{Barnes:2010, Barnes_Kim:2010}.

For detailed modeling of specific stars, we provide an interface between the YaPSI tracks and an open-source, freely available \texttt{Fortran} code that automatically performs a best-fitting search of the observed stellar parameters using a Monte Carlo Markov-Chain approach.

Finally, the YaPSI models can be useful in the classical isochrone fitting of the color-magnitude diagrams (CMDs) of open and globular stellar clusters (see Section \ref{ssec_clusters}). 
Notably, the chemical compositions available in YaPSI include, in addition to a typical span of metallicities, a broader range of helium abundances than is usually considered; moreover, these two quantities are assigned independently of each other, i.e., the full range of helium abundances is available at each metallicity.  
This is in contrast with the more widespread choice in the literature, where helium content is constrained to remain in lockstep with metallicity through a fixed $\Delta Y/ \Delta Z$ relation, assigned on the basis of a Galactic nucleosynthesis model \citep[e.g.,][]{Yi_ea:2003}.
The plausibility of stellar helium abundances being locally uncorrelated with metallicity has been discussed in the literature in the past in several contexts (see, e.g., \citealt{Demarque_McClure:1977}, and more recently \citealt{Norris:2004} and \citealt{Lee_ea:2005}).  

The choice of independent metallicity and helium abundance parameters provides not only additional flexibility in stellar population studies, but also the ability to explore binaries and multiple stars that may have undergone enrichment through interaction during their evolution.

Together with the YaPSI isochrones, we provide a \texttt{Fortran} routine for interpolation in age, metallicity, and helium content, similar to the code available for the YY isochrones.

This paper is organized as follows. 
The input physics included in the models is discussed in Section \ref{sec_input}.  
The parameters of the grid and the stellar evolution tracks are described in Section \ref{sec_tracks}.
The isochrones and their newly-implemented construction procedure are presented in Section \ref{sec_isochrones}.  
In Section \ref{sec_observations} we compare the YaPSI mass-luminosity and mass-radius relations with the best observations available, and we discuss the isochrone fitting of the color-magnitude diagrams of selected open and globular stellar clusters.

\section{Input physics}
\label{sec_input}

\subsection{Basic input physics and parameters}

The models in the grid have been constructed using the Yale Rotational stellar Evolution Code (YREC) in its non-rotational configuration \citep{Demarque_ea:2008}.
The details of the microphysics used are as follows.

The code uses the OPAL Rosseland mean opacities in the temperature range $\log T \geq 4.5$ \citep{Rogers_Iglesias:1995,Iglesias_Rogers:1996}, ramped with the ``low temperature" opacities of \citet{Ferguson_ea:2005} for $4.4 \leq \log T \leq 4.5$.
Conductive opacities corrections are introduced for $\log T \geq 4.2$ and $\log \rho \geq 2 \log T - 13$, using an analytical fit based on the work of \citet{Hubbard_Lampe:1969} and \citet{Canuto:1970}.
We adopt the \citet{Grevesse_Sauval:1998} value of the solar metallicity, $(Z/X)_\odot = 0.0230$.

The treatment of convection is based on the mixing length theory \citep{BV58}, according to the formulation of \citet{Paczynski:1969}.

\begin{figure}
\begin{center}
\includegraphics[width=0.49\textwidth]{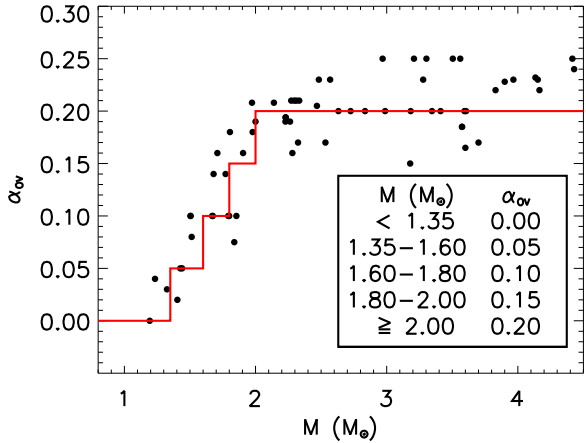}
\caption{{Mass dependence of the core overshoot parameter implemented in YaPSI. The semi-empirical determinations of $\alpha_{\rm OV}$ by \citet{Claret_Torres:2016} (black points) are plotted together with our ramping function (red line).}}
\label{coreovsh}
\end{center}
\end{figure}

The extent of convective core overshoot is assumed to be proportional to the local pressure scale height $H_P$ at the edge of the core (as defined by the Schwarzschild criterion), scaled by the factor $\alpha_{\rm OV}$.  
The pressure scale height, however, approaches infinity at the stellar center, and it becomes physically unrealistic as a measure of overshoot length when the convective core is very small.   
As a solution, our code uses the simple recipe suggested by \citet{Wuchterl_Feuchtinger:1998}, which limits the core overshoot length scale to a fraction of the geometrical distance to the stellar center.
We assign $\alpha_{\rm OV}$ according to the recent work of \citet{Claret_Torres:2016}.
These authors have determined the best-fitting value of the core overshoot parameter for a sample of accurately characterized stars, strategically distributed across the HR diagram to constrain its mass dependence.
Although the sample spans the range $-1 \lesssim$ [Fe/H] $\lesssim 0$, no dependence of $\alpha_{\rm OV}$ on metallicity was found.
We have constructed a ramping function to fit the semi-empirical determinations of $\alpha_{\rm OV}$ by \citet{Claret_Torres:2016}; see Figure \ref{coreovsh}.
This can be contrasted with the approach in Paper IV of YY \citep{Demarque_ea:2004}, in which an estimate to the value of $\alpha_{\rm OV}$ is taken to match the observed CMDs of stellar clusters.   
Other such approximations have been introduced by \citet{Pietrinferni_ea:2004,Bressan_ea:2012,Mowlavi_ea:2012}.
For another approach to core overshoot, based on a diffusive treatment, see \citet{Heger_ea:2000, Moravveji_ea:2015}.

The standard choices of the Equation of State (EOS) and of the surface boundary conditions in YREC are the OPAL 2005 EOS \citep[][]{Rogers_Nayfonov:2002} and the classical Eddington gray $T$--$\tau$ relationship, respectively. 
In order to meet the specific input physics requirements of low-mass models \citep{Allard_ea:1997, Chabrier_Baraffe:2000, Sills_ea:2000}, for tracks of mass $M_*<0.6\, M_\odot$, the SCVH EOS  \citep{Saumon_ea:1995} is ramped with the OPAL 2005 EOS in regions where $5000$ K $\leq T \leq 6000$ K. 
Moreover, for the same mass range, surface boundary conditions providing the photospheric pressure in tabular form, derived from \texttt{PHOENIX} atmospheric models \citep{Hauschildt_ea:1999}, are used in place of the gray Eddington $T$--$\tau$ relation.
These tables are based on the ``BT-Settl" batch of \texttt{PHOENIX} model atmospheres \citep{Allard_ea:2011}\footnote{\texttt{https://phoenix.ens-lyon.fr/Grids/BT-Settl/}. As discussed in \cite{Spada_ea:2013}, this introduces a small inconsistency with the solar mixture assumed in our interior calculation.}.

Both helium and heavy elements diffusion are implemented in our models, based on Loeb's formulation \citep{Bahcall_Loeb:1990, Thoul_ea:1994}.
No extra mixing to limit the effect of diffusion is included (as required to match the observed surface abundances of population II stars: e.g., \citealt{Richard_ea:2002}; see also \citealt{Gruyters_ea:2013}).
Mass loss is not taken into account.

The nuclear energy generation rates are those recommended by \citet{Adelberger_ea:2011}. 
As noted by \citet{Maxted_ea:2015}, the most significant difference with the rates of \citet{Bahcall_Pinsonneault:1992}, adopted in the YY calculations, is for the reaction $^{14}N(p,\gamma)^{15}O$, which has been revised by a factor of $\approx 2$.

\subsection{Mass-dependent input physics}
\label{ssec_massdep}

The EOS and the surface boundary conditions depend on the stellar mass as follows:
\begin{itemize}
\item {\it ``Low-mass configuration"}, OPAL$+$SCVH EOS, \texttt{PHOENIX}--derived surface boundary conditions: for tracks in the mass range $0.15 \leq M/M_\odot \leq 1.10$;
\item {\it ``Standard configuration"}, OPAL EOS, Eddington gray surface boundary conditions: for tracks in the range $0.60 \leq M/M_\odot \leq 5.00$.
\end{itemize}
The two subsets overlap between $0.6\, M_\odot$ and $1.1\, M_\odot$, to allow for intercomparison of the results.
At given mass and chemical composition, the evolutionary track is most affected by the different EOS and atmospheric boundary conditions choices during the pre-main sequence contraction and the red giant phase.
A smooth transition between the two subsets occurs at $\approx 0.9 \, M_\odot$ (this threshold is moderately dependent on chemical composition; see Appendix \ref{overlap} for more details).

\begin{figure}
\begin{center}
\includegraphics[width=0.49\textwidth]{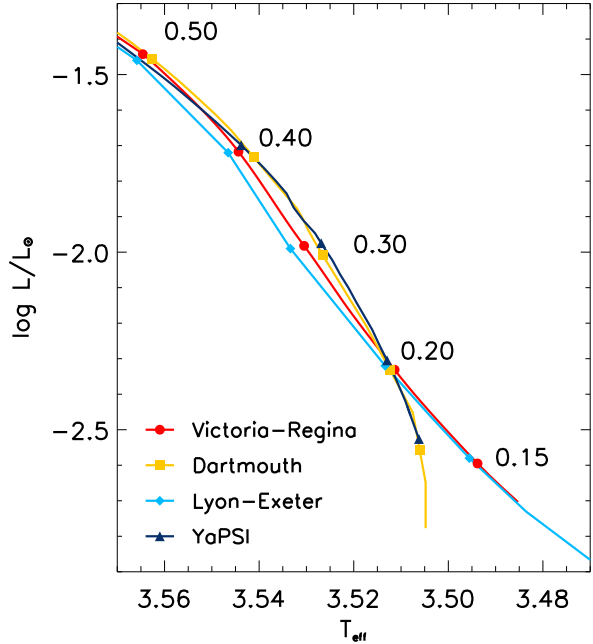}
\caption{{Comparison of the low-mass end of the $5$ Gyr, solar composition YaPSI isochrone with its counterpart from other recent works. The Victoria-Regina \citep{Vandenberg_ea:2014} and the Lyon-Exeter models \citep{Baraffe_ea:2015} implement non-gray atmospheric boundary conditions attached at $\tau=100$, while the Dartmouth ones \citep{Dotter_ea:2008} share with YaPSI the choice of a matching point at $T=T_{\rm eff.}$}}
\label{psurf}
\end{center}
\end{figure}

It should be noted that, when using surface boundary conditions derived from detailed model atmospheres, such as the \texttt{PHOENIX} Bt-Settl ones, the atmospheric pressure must be attached to the interior model at a suitable location.
Commonly adopted choices include the layer where $T=T_{\rm eff}$, or a deeper location, specified in terms of optical depth (e.g., $\tau=100$). 
Although the two strategies yield results in good agreement with each other for masses larger than $\approx 0.2 \, M_\odot$, below this threshold the latter alternative results in systematically lower $T_{\rm eff}$ at a given mass, and should be preferred on theoretical grounds \citep[e.g.,][]{Chabrier_Baraffe:2000}.
This effect is illustrated in Figure \ref{psurf}, where we compare the low-mass end of our $5$ Gyr, solar composition isochrone with its counterpart from other isochrone sets available in the literature. 
In our calculation, we have adopted the $T=T_{\rm eff}$ matching point \citep[see also][]{Sills_ea:2000,Spada_ea:2013}.
In the Figure, this effect is clearly visible in the portion of the isochrone corresponding to masses below $0.2 \, M_\odot$.
At a given mass, the effect on the luminosity is much smaller in comparison with that on the effective temperature (which is $\approx 80$ K at  $0.15 \, M_\odot$).

\section{Evolutionary tracks}
\label{sec_tracks}

\subsection{Mass and chemical composition ranges}

The grid covers the mass range $0.15 \leq M/M_\odot \leq 5.00$ for each chemical composition specified by the parameters $(X_0,Z_0)$ in Table~\ref{xzcomp}.  
To optimize the number of tracks to be calculated, ensuring at the same time a dense coverage of the mass range, the mass increments vary as follows: between $0.15$ and $0.40 \, M_\odot$: $\Delta M = 0.01\, M_\odot$; $0.40$--$0.90 \, M_\odot$: $\Delta M = 0.02\, M_\odot$; $0.90$--$1.80 \, M_\odot$: $\Delta M = 0.05\, M_\odot$; $1.80$--$3.00 \, M_\odot$: $\Delta M = 0.10\, M_\odot$; $3.00$--$5.00 \, M_\odot$: $\Delta M = 0.20\, M_\odot$.

The metallicity range spanned by the grid goes from moderately metal-poor to moderately metal-rich: [Fe/H]$_0=-1.5, \, -1.0, \, -0.5, \, 0.0, \, +0.3$ (with respect to the \citealt{Grevesse_Sauval:1998} solar mixture).
The abundances of individual elements are solar-scaled, i.e., $\alpha$-elements enhancement is ignored.
For each value of [Fe/H]$_0$, models with the following initial helium mass fractions are available: $Y_0 = 0.25, \, 0.28, \, 0.31, \, 0.34, \, 0.37$.
Note that, in this way, $Y_0$ and [Fe/H]$_0$ are treated as independent parameters; in other words, no fixed enrichment relation $\Delta Y/\Delta Z$ is assumed.

The initial chemical compositions available in the grid are listed in Table \ref{xzcomp}. 
Note that, due to the effect of the diffusion of metals, [Fe/H] is not constant during the evolution, the change being most pronounced for the late evolutionary stages of the long-lived low-mass stars and at metal-poor compositions.

\begin{table}
\caption{Initial chemical compositions.}
\begin{center}
\begin{tabular}{cccc}
\hline
\hline
[Fe/H] & $Y_0$ & $X_0$ & $Z_0$ \\
\hline
\multirow{5}{*}{-1.5} & 0.25 &  0.749455 &  0.000545 \\
  & 0.28 & 0.719477 & 0.000523  \\
  & 0.31 & 0.689499 & 0.000501  \\
  & 0.34 & 0.659520 & 0.000480  \\
  & 0.37 & 0.629542 & 0.000458  \\
\hline
\multirow{5}{*}{-1.0} & 0.25 & 0.748279  & 0.001721  \\
  & 0.28 & 0.718348 & 0.001652  \\
  & 0.31 & 0.688417 & 0.001583  \\
  & 0.34 & 0.658485 & 0.001515  \\
  & 0.37 & 0.628554 & 0.001446  \\
\hline
\multirow{5}{*}{-0.5} & 0.25 & 0.744584 & 0.005416  \\
  & 0.28 & 0.714801 & 0.005199  \\
  & 0.31 & 0.685018 & 0.004982  \\
  & 0.34 & 0.655234 & 0.004766  \\
  & 0.37 & 0.625451 & 0.004549  \\
\hline
\multirow{5}{*}{\phantom{-}0.0} & 0.25 & 0.733138 & 0.016862 \\
  & {0.28} & {0.703812} & {0.016188} \\
  & 0.31 & 0.674487 & 0.015513  \\
  & 0.34 & 0.645161 & 0.014839  \\
  & 0.37 & 0.615836 & 0.014164  \\
\hline
\multirow{5}{*}{\phantom{-}0.3} & 0.25 & 0.717092 & 0.032908  \\
  & 0.28 & 0.688408 & 0.031592  \\
  & 0.31 & 0.659725 & 0.030275  \\
  & 0.34 & 0.631041 & 0.028959  \\
  & 0.37 & 0.602357 & 0.027643  \\
\hline
\end{tabular}
\end{center}
Note. -- Due to the effect of metal diffusion, [Fe/H] is not constant during the evolution. 
\label{xzcomp}
\end{table}

All the tracks were constructed using a solar-calibrated value of the mixing length parameter, $\alpha_{\rm MLT}$.
We performed a standard solar model calibration \citep[e.g.,][]{Basu_Antia:2008} for both choices of surface boundary conditions used in the grid.
 The adopted values of the solar parameters are: $M_\odot=1.988\cdot 10^{33}$ g, $L_\odot=3.828\cdot 10^{33}$ erg s$^{-1}$, $R_\odot=6.957\cdot 10^{10}$ cm, as recommended by the XXIXth IAU General Assembly resolution B3\footnote{\url{https://www.iau.org/static/resolutions/IAU2015_English.pdf}}. 
The best solar models ($\log L/L_\odot \lesssim 10^{-7}$, $\log R/R_\odot \lesssim 10^{-7}$, $(Z/X)_\odot - 0.023 \lesssim 10^{-6}$) are obtained with $\alpha_{\rm MLT}=1.91804$, $Y_0 = 0.277486$ for the \texttt{PHOENIX} surface boundary conditions, and with $\alpha_{\rm MLT}=1.82126$; $Y_0 = 0.277550$ when using the Eddington $T$-$\tau$ relation.
Due to the effects of helium and metal diffusion, the initial composition of the Sun in our calibrated solar models corresponds to $(Z_0/X_0)=0.0267$, or [Fe/H]$_0=0.065$.

\begin{figure*}
\begin{center}
\includegraphics[width=0.9\textwidth]{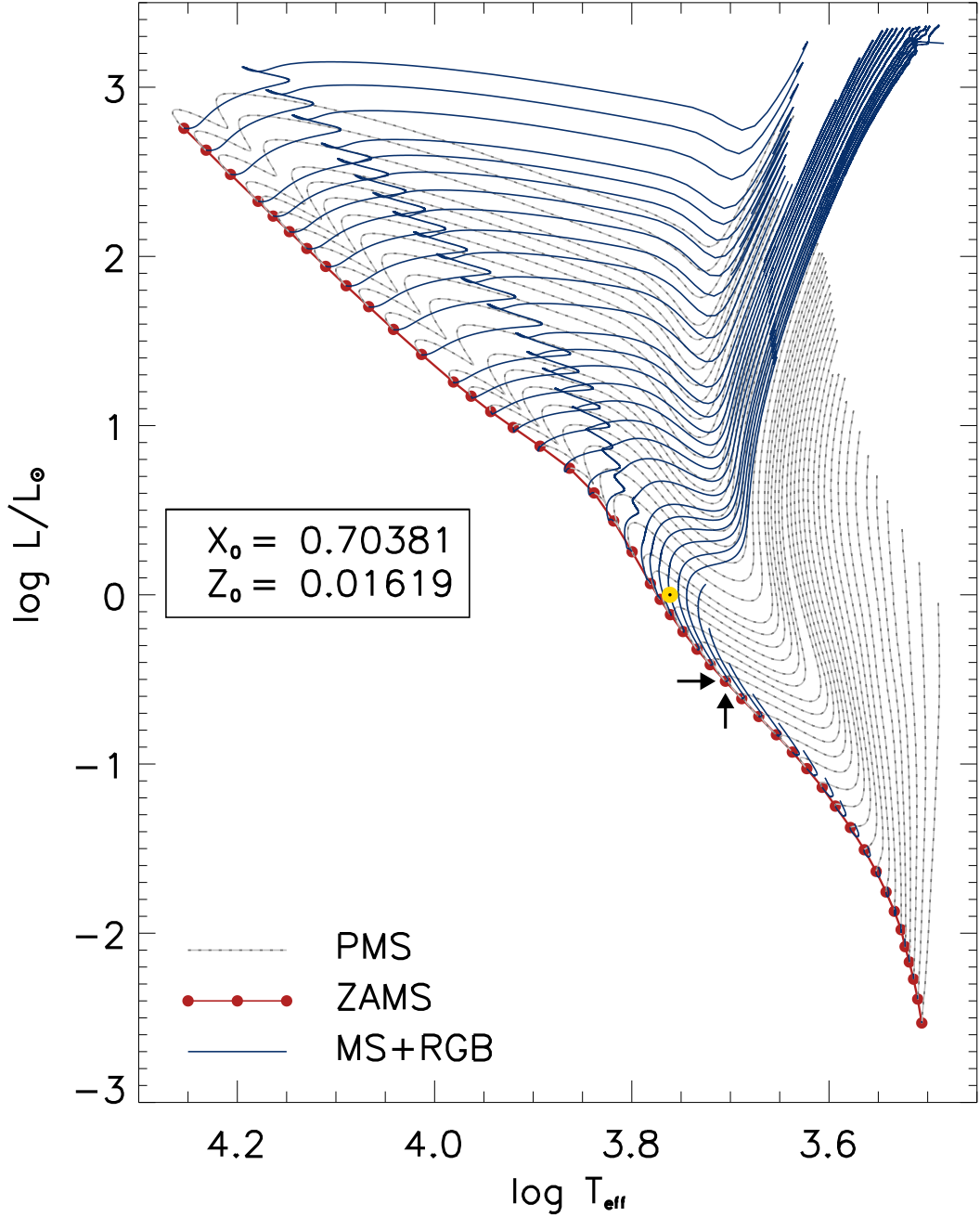}
\caption{A selection of the YaPSI evolutionary tracks with initial composition close to solar ($X_0 = 0.7038$, $Z_0 = 0.0162$).  
Tracks plotted: $0.15$--$0.82 \, M_\odot$, constructed using the low-mass input physics configuration (OPAL05+SCVH EOS, \texttt{PHOENIX} ``BT-Settl" atmospheres, see section \ref{ssec_massdep}); $0.86$--$5.00 \, M_\odot$, constructed with the standard input physics configuration (OPAL05 EOS, Eddington gray atmosphere); the arrows highlight the ZAMS locus of the $0.82 \, M_\odot$ track, where the transition occurs. The Sun is also shown for reference. The evolution of low-mass tracks is only plotted up to $15$ Gyr for clarity.}
\label{tracks}
\end{center}
\end{figure*}

The tracks cover the evolution through the pre-main sequence and the main sequence, and the post-main sequence up to helium flash or central helium burning ignition (depending on the mass). 
Note that, at the lowest masses, the tracks extend well beyond the current age of the Universe of $\approx 13.7$ Gyr \citep[for a discussion of this interesting regime in connection with the theory of red giant evolution, see][]{Laughlin_ea:1997}. 
The evolutionary calculations start from initial models with homogeneous composition (with \citealt{Grevesse_Sauval:1998} solar mixture) and polytropic structure. 
They are similar to the initial models used in the YY project \citep{Yi_ea:2001} and in the early YaPSI release \citep{Spada_ea:2013}.  
Some evolutionary tracks for the chemical composition closest to solar ($X_0 = 0.7038$, $Z_0 = 0.0162$), implementing low-mass and standard input physics, are shown in Figure \ref{tracks}. 

\subsection{Best-fitting of individual stars with YaPSI}

Based on the YaPSI tracks, we have constructed input files compatible with \texttt{BAGEMASS}, a freely available\footnote{\url{https://sourceforge.net/projects/bagemass/}}, open source Fortran code developed by \citet{Maxted_ea:2015}, that implements a best-fitting search of the observed parameters of a star using a Monte Carlo Markov-Chain approach.  
This code calculates the posterior probability distribution for mass, age, and initial metallicity of a star from its observed mean density, effective temperature, luminosity, and surface metallicity, given a set of theoretical evolutionary tracks.
Element diffusion is taken into account in the optimization process.

We provide YaPSI input files ready to be ingested into \texttt{BAGEMASS}, only requiring trivial modifications in the first lines of the code to be used.
To keep our input files as close as possible to the originals, we have adopted the same procedure used by \citet{Maxted_ea:2015}: the interpolation of the YaPSI tracks has been performed using the bi- and tri-cubic spline algorithms implemented in the \texttt{PSPLINE} package\footnote{\url{http://w3.pppl.gov/ntcc/PSPLINE}}, and the resulting interpolated grids have been bundled together as .fits format files using the \texttt{FITSIO} library (\citealt{Pence:1998}; see also \citealt{Maxted_ea:2015} for details).

\section{Isochrones}
\label{sec_isochrones}

We have constructed a set of isochrones between $1$ Myr and $20$ Gyr for each of the initial compositions listed in Table~\ref{xzcomp}.
Note that the choice of ages coincides with that of the YY isochrones, in order to facilitate comparisons and ensure retro-compatibility of user-generated codes. 

\begin{figure*}
\begin{center}
\includegraphics[width=0.9\textwidth]{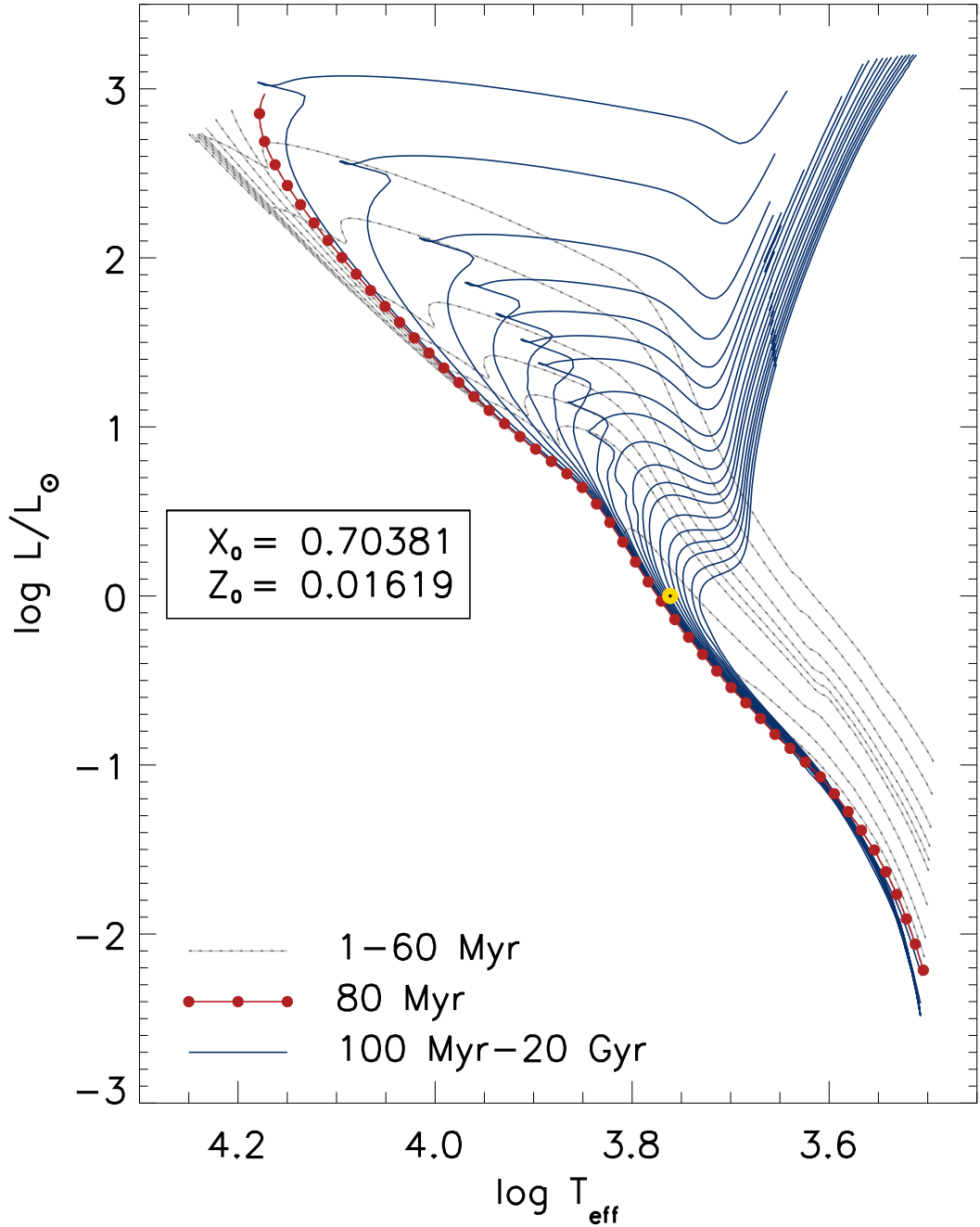}
\caption{YaPSI isochrones for composition close to solar ($X_0 = 0.7038$, $Z_0 = 0.0162$).
The $80$ Myr isochrone is a reasonable approximation of the ZAMS across the mass range of YaPSI; the position of the Sun is marked for reference. In sufficiently old isochrones (i.e., age $\gtrsim1$ Gyr) the ``luminosity bump" \citep{Thomas:1967} is visible on the red giant branch at $\log L/L_\odot \approx 1.5$. To enhance the clarity of the plot, only a subset of the isochrones that reach the tip of the red giant branch is shown.}
\label{iso}
\end{center}
\end{figure*}

To construct the isochrones, the low-mass configuration tracks were used up to $M_* < 0.6 \, M_\odot$ and the standard configuration tracks for $M_* > 1.1\, M_\odot$, while in the overlap region, the switch from low-mass to standard tracks was performed at a composition-dependent mass (usually between $0.7$ and $0.9\, M_\odot$, see the discussion in Appendix \ref{overlap}).
In this way, we ensure the smoothest transition possible without recourse to post-processing of the tracks (for an alternative approach, see \citealt{Choi_ea:2016}).
The YaPSI isochrones for $X_0 = 0.7038$, $Z_0 = 0.0162$ are plotted in Figure \ref{iso}.

\subsection{Isochrone construction procedure}

The procedure for isochrone construction used in this work has been coded from scratch in order to be specifically tailored to the YaPSI evolutionary tracks.
It is based on the method discussed by \citet{Prather:1976}.
Only a brief description is given here; for more details on the subject of isochrone construction, see also the recent review by \citet{Dotter:2016} and references therein.

The most obvious procedure to construct an isochrone is to simply interpolate a set of tracks to the desired age.
Although this approach may be viable up to the mid-main sequence, it will provide a poor representation of the fast-paced evolutionary phases, such as the late main sequence and beyond \citep[see the discussion in][]{Dotter:2016}.

A much more reliable method is based on re-mapping each track as a function of uniformly spaced ``Equivalent Evolutionary Points" (EEPs) before interpolation.
This is accomplished by, firstly, locating on each track a number of primary EEPs. 
As suggested by \citet{Prather:1976}, useful primary EEPs can be defined as the evolutionary stages at which a certain fixed value of, e.g., central hydrogen content, $X_c$ (on the main sequence), or helium core mass, $M_c$ (during the post-main sequence), is first attained. 
Each portion of the track between two primary EEPs is then further subdivided into secondary EEPs.

The secondary EEPs are uniformly distributed with respect to an intrinsic arc length coordinate $s$ defined on the track.
In view of the evolution of a star in the theoretical HR diagram, \citet{Prather:1976} proposed the following definition of $\Delta s$ between two adjacent EEPs:
\begin{equation}
(\Delta s)^2 \equiv (a\, \Delta \log L/L_\odot)^2 + (b\, \Delta \log T_{\rm eff})^2 + (c\, \Delta \log \tau)^2,
\end{equation}
where $\tau$ is the age and the $\Delta$'s represent the change of the respective variables between the two EEPs in question; $(a,b,c)$ are opportunely chosen weights. 
In our experience, the following choices have proved satisfactory: pre-main sequence: $(a,b,c)=(1,10,1)$; main sequence $(a,b,c) = (1,10,0)$; red giant branch: $(a,b,c)=(1,0,0)$.
The definition of $\Delta s$ ensures that stellar parameters at the same EEP can be meaningfully compared between tracks of different mass. 

Once the uniformly EEP-spaced tracks have been constructed, we proceed as follows:
\begin{enumerate}
\item For a given EEP, locate the pair of tracks in the set whose ages $\tau_1$ and $\tau_2$ at that EEP bound the desired isochrone age $\tau_{\rm iso}$;
\item Interpolate (e.g., linearly) in mass between the two tracks:
\begin{eqnarray*}
f &=& \frac{\log \tau_{\rm iso} - \log \tau_1}{\log \tau_2 - \log \tau_1}; 
\\  
m_{\rm iso} &=& m_1 + f\, (m_2-m_1),
\end{eqnarray*}
where $m_1$ and $m_2$ are the masses of the two tracks, found in the previous step, for which $\tau_1 < \tau_{\rm iso} < \tau_2$;
\item Find the effective temperature and the luminosity corresponding to $\tau_{\rm iso}$ from the two tracks:
\begin{eqnarray*}
\left(\log {L}/{L_\odot} \right)_{\rm iso} &=& \left(\log {L}/{L_\odot} \right)_1 
\\
&+& f\, \left[ \left( \log {L}/{L_\odot} \right)_2 - \left(\log {L}/{L_\odot} \right)_1 \right],
\\
\left(\log T_{\rm eff} \right)_{\rm iso} &=& \left(\log T_{\rm eff}  \right)_1 
\\
&+& f\, \left[ \left(\log T_{\rm eff}  \right)_2 - \left(\log T_{\rm eff}  \right)_1 \right];
\end{eqnarray*}
\item Loop over all the EEPs and all the desired isochrone ages.
\end{enumerate}

\begin{figure*}
\begin{center}
\includegraphics[width=0.85\textwidth]{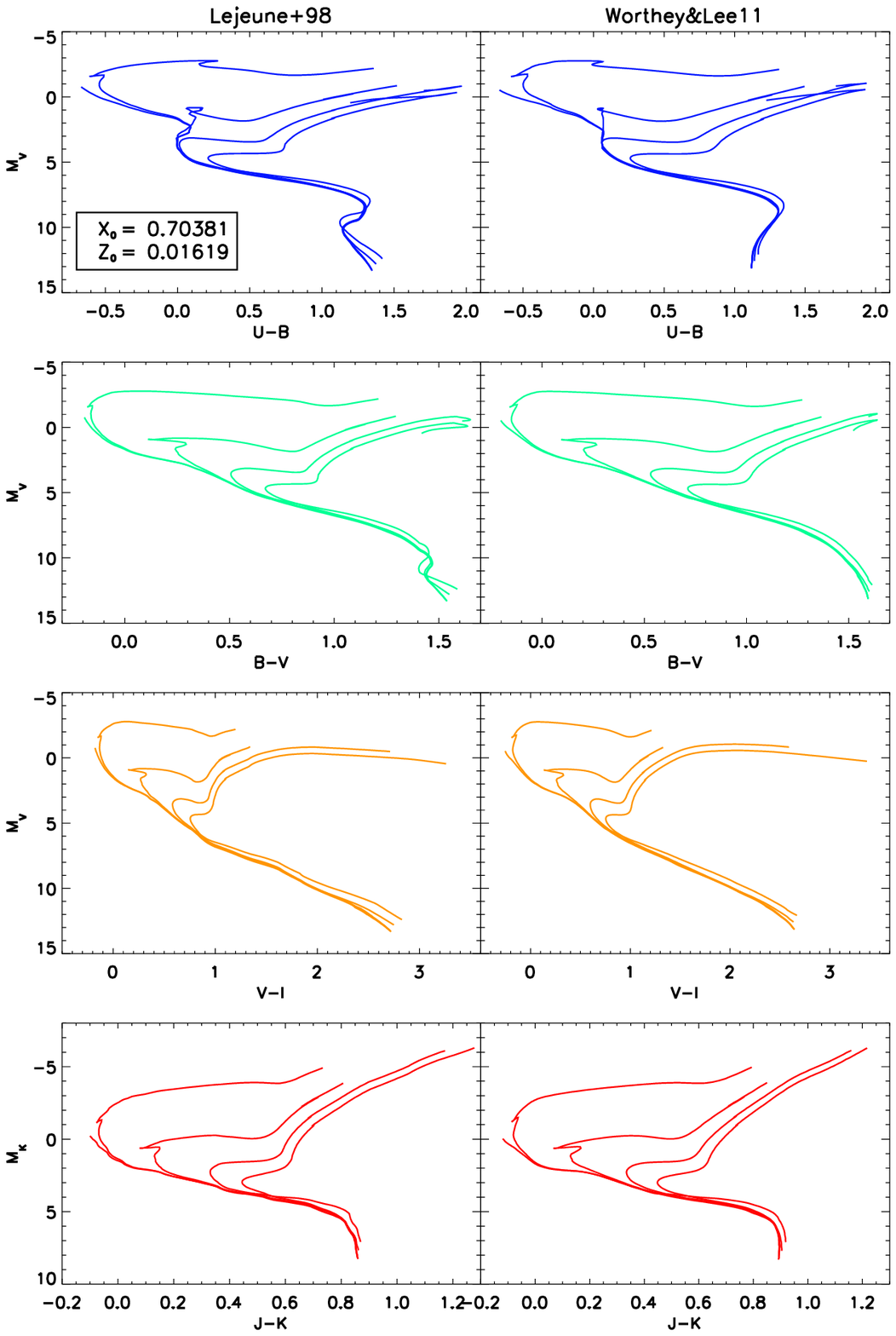}
\caption{{YaPSI isochrones (ages $=0.04$, $0.1$, $1$, $5$, and $15$ Gyr) of composition close to solar ($X_0 = 0.7038$, $Z_0 = 0.0162$) in various color-magnitude diagrams, obtained with the \citet{Lejeune_ea:1997,Lejeune_ea:1998} color transformations (left column), and the \citet{Worthey_Lee:2011} ones (on the right).}}
\label{iso_col}
\end{center}
\end{figure*}

\subsection{Semi-empirical $T_{\rm eff}$--color transformations}

Along with the theoretical parameters $\log L/L_\odot$, $\log T_{\rm eff}$, $\log g$, observational quantities such as magnitudes and colors are provided for each isochrone.
To perform the transformation to magnitudes and colors, we have used two different prescriptions, both semi-empirical, from the literature: that of \citet{Lejeune_ea:1997,Lejeune_ea:1998}, which was also used in the YY isochrones, and the more recent one of \citet{Worthey_Lee:2011}.
Both color transformations are provided in tabular form, from which the colors corresponding to the desired values of [Fe/H], $T_{\rm eff}$, and $\log g$ can be obtained by interpolation.
We have used the ``BaSeL 2.2 SED-corrected" version of the \citet{Lejeune_ea:1997,Lejeune_ea:1998} tables, available from the BaSeL interactive server\footnote{http://www.astro.mat.uc.pt/BaSeL2/}.
Trilinear interpolation in [Fe/H], $T_{\rm eff}$, and $\log g$ was used.
For the \citet{Worthey_Lee:2011} color calibration, user-friendly interpolation tools are provided along with the look-up tables.
Their code uses bilinear interpolation in [Fe/H] and $\log g$, and quartic interpolation in $T_{\rm eff}$ (based on the polynomial interpolation algorithm \texttt{polint} from \citealt{Press:1992}). 

The variation of [Fe/H] due to metal diffusion is taken into account in the color transformations. 
This is, in general, a small effect, but can be non-negligible for the long-lived low-mass models and/or at low metallicity.

YaPSI isochrones are available with color calibrations performed according to both the \citet{Lejeune_ea:1997,Lejeune_ea:1998} and the \citet{Worthey_Lee:2011} recipes (see, for example, Figure \ref{iso_col}). 
In both cases, we provide the absolute $V$ magnitude, $M_V$, the $UBVRI$ colors in the Johnson-Cousins system, and the $JHK$ colors in the homogeneized \citet{Bessell_Brett:1988} system.
Sample comparisons between the two $T_{\rm eff}$--color calibrations available in YaPSI can be found in Section \ref{ssec_clusters}.

\section{Comparison with observations}
\label{sec_observations}

In this Section we compare the YaPSI models with selected observations.
In the first two subsections we discuss the mass-luminosity and mass-radius relations, respectively.

The theoretical mass-luminosity relations derived from $1$ Gyr YaPSI isochrones are compared with the sample of nearby red dwarf astrometric binaries studied by \citet{Benedict_ea:2016}.

The mass-radius relation is best constrained by double-lined eclipsing binaries, for which both the radius and the mass of each component can be derived with an accuracy of a few percent or better \citep{Torres_ea:2010, Southworth:2015}; at this level of accuracy, the ``inflation" of observed stellar radii compared to the model predictions becomes apparent.

We conclude this Section with the isochrone fitting of some well-studied stellar clusters.

\subsection{The mass--luminosity relation of M dwarf stars}
\label{ssec_benedict}

\cite{Benedict_ea:2016} have recently published empirical MLRs for low-mass stars ($0.08$--$0.62\, M_\odot$) of exquisite precision. 
Their $V$-band and $K$-band MLRs are based on a sample of $47$ nearby red dwarf stars, whose dynamical masses are determined with an accuracy of $4$ percent or better (primarily using astrometric data from the Fine Guidance Sensors of the {\it Hubble Space Telescope}; see their paper for details on the orbit reconstruction process). 

The availability of such empirical MLRs is a major milestone in the physics of low-mass stars for their accuracy, and general applicability (as well as a remarkable technical achievement, for the difficulty of the task and the long-term commitment required to complete it).
Such relations are of crucial importance for observers, to convert observed luminosities into masses, as well as for theorists, to test the numerous assumptions that enter stellar models \citep{Andersen:1991}. 

\begin{figure}
\begin{center}
\includegraphics[width=0.49\textwidth]{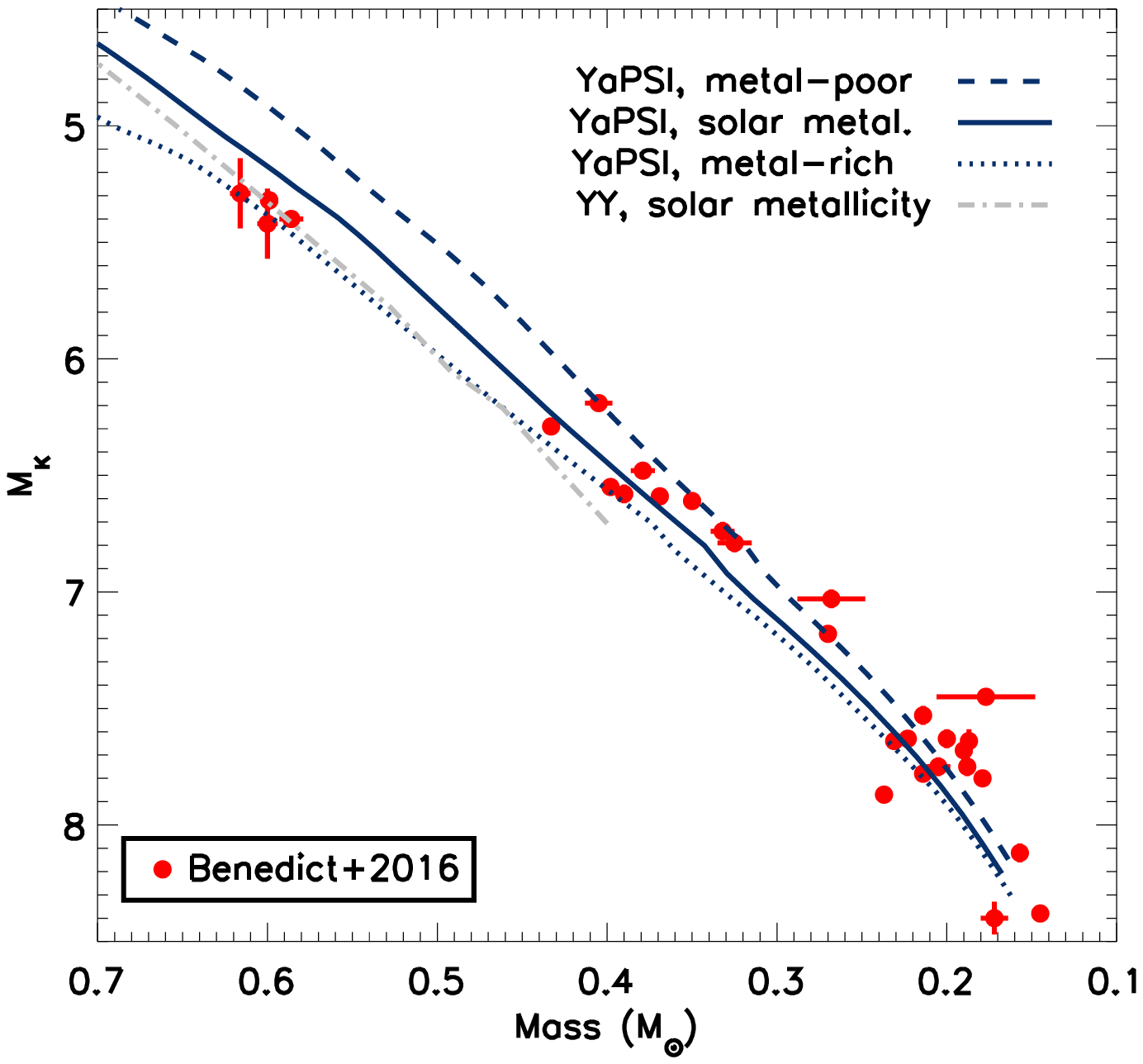}
\includegraphics[width=0.49\textwidth]{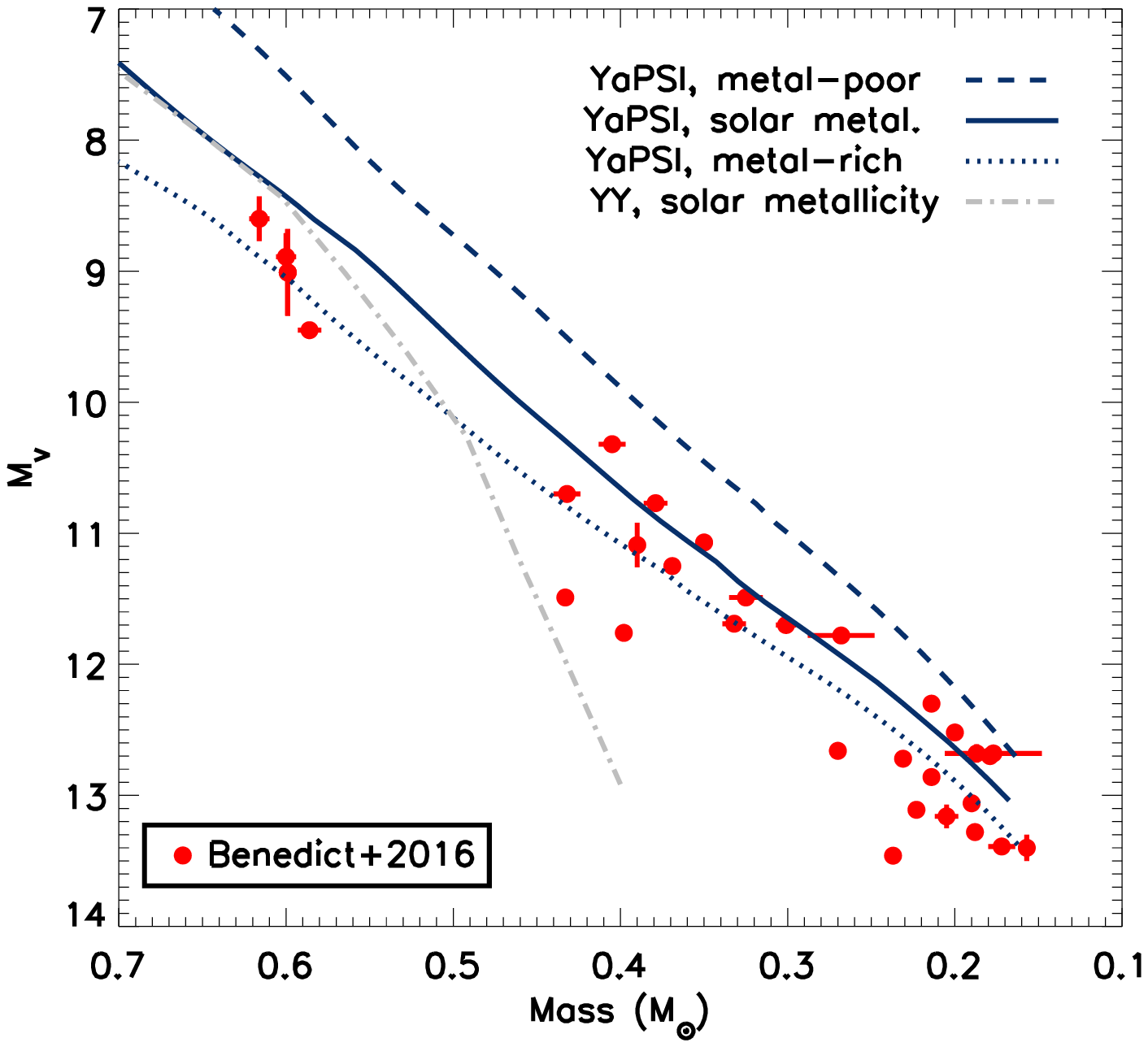}
\caption{Comparison of the YaPSI mass--luminosity relations with their empirical counterparts in the $K$ and $V$ band from \citet{Benedict_ea:2016}. 
The data points are plotted as red dots (note that the error bars are mostly within the size of the dots). 
A very good agreement is found for the mass--$M_K$ relations, while the predicted mass-$M_V$ relations are slightly overluminous with respect to the data.}
\label{mlr}
\end{center}
\end{figure}

We compare the YaPSI theoretical mass--$M_V$ and mass--$M_K$ relations with the empirical relations of \citet{Benedict_ea:2016} in Figure \ref{mlr}.
The theoretical MLRs in the Figure are obtained with the \citet{Worthey_Lee:2011} color transformations; the \citet{Lejeune_ea:1997,Lejeune_ea:1998} transformations yield a comparable level of agreement.

Solar metallicity, metal-poor ([Fe/H]$_0=-0.5$), and metal-rich ([Fe/H]$_0=+0.3$) theoretical MLRs at $1$ Gyr are plotted in the Figure; solar metallicity YY isochrones are also shown for comparison.
Any isochrone between $1$ and $10$ Gyr results in theoretical MLRs essentially equivalent to the ones shown.
This is a consequence of the very slow-paced evolution of stars in this mass range. 

The metallicity dependence of the K-band MLR is much weaker than that of the V-band MLR, especially for $M_*\lesssim 0.4\, M_\odot$. 
As discussed in detail by \citet{Baraffe_ea:1998}, this is the result of two competing effects (the TiO and VO oxygen molecules formation at low $T_{\rm eff}$, and the metallicity dependence of $T_{\rm eff}$) compensating and reinforcing each other in the K- and V-band, respectively.

The agreement of the YaPSI isochrones is clearly better than that of the YY ones below approximately $0.6\, M_\odot$.
This is exactly as expected, since the improved input physics of the ``low-mass configuration" is especially important for stars in this mass range \citep[see, e.g.,][]{Allard_ea:1997,Chabrier_Baraffe:2000}. 
The YY and the solar YaPSI relations are very similar above $\approx 0.6\, M_\odot$.

The YaPSI mass--$M_K$ relation at solar composition is in excellent agreement with the empirical one. 
For stars more massive than about $0.25 \, M_\odot$, the scatter of the data points is also comparable with the gap between the [Fe/H]$_0=-0.5$ and [Fe/H]$_0=+0.3$ isochrones (this metallicity range is representative of that spanned by the stars in the \citealt{Benedict_ea:2016} sample; see their table $14$).   
The theoretical mass--$M_V$ relation is slightly overluminous in comparison with the data.
The size of the spread between the metal-poor and the metal-rich isochrones is comparable to the observed scatter.
In the mass range covered by the YaPSI models, the slope of the MLRs is not significantly affected by the details of the implementation of the atmospheric boundary conditions (cf. Figure \ref{psurf} and the discussion at the end of Section \ref{sec_input}).

\subsection{The mass--radius relation}
\label{ssec_rinfl}

The most stringent test of the theoretical mass--radius relation is provided by double-lined eclipsing binaries (DEBs), for which both the mass and the radius can be derived simultaneously with an accuracy of a few percent.
A long-standing issue in the theory of low-mass stellar objects ($M_* \lesssim 0.75\, M_\odot$) is the discrepancy between observed and modeled radii and surface temperatures \citep[e.g.,][]{Hoxie:1973,Lacy:1977,Lopez-Morales:2007,Torres_ea:2010}. 
Theoretical stellar radii are typically underestimated by $3$--$5\%$ and effective temperatures are overestimated by $5$--$10\%$ with respect to the observations \citep[e.g.,][]{Torres_ea:2010,Feiden_Chaboyer:2012a,Spada_ea:2013}.

\begin{figure}
\begin{center}
\includegraphics[width=0.49\textwidth]{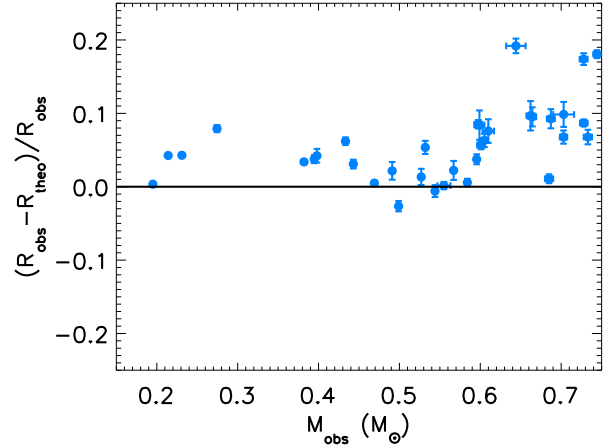}
\caption{YaPSI mass--radius relation compared with high-accuracy DEBs measurements.
The residuals shown are calculated with respect to a solar metallicity, $1$ Gyr isochrone for the stars in the DEBCAT sample \citep{Southworth:2015}. }
\label{massradius}
\end{center}
\end{figure}

This is illustrated in Figure \ref{massradius}, where the residuals calculated from a YaPSI solar metallicity, $1$ Gyr isochrone are shown for the stars in the DEBCAT sample \citep{Southworth:2015}.
Only low-mass stars ($M_*\lesssim 0.75\, M_\odot$), whose evolution from the ZAMS within a Hubble time is very modest, are shown, thus making the comparison with a single isochrone meaningful.
The sample contains, with few exceptions, only stars whose parameters are known with $3\%$ accuracy, or better \citep[see][]{Southworth:2015}.
In the Figure, there is a clear excess of stars with observed radii larger than the theoretical prediction (i.e., an excess of positive residuals). 
This illustrates the radius inflation problem.

As a further test of the mass--radius relation, as well as to demonstrate the use of the \texttt{BAGEMASS} tool in combination with the YaPSI input files, we have run the Monte Carlo Markov-Chain (MCMC) best-fit search on the two samples of DEBs and planet-host stars listed in tables 2 and 3, respectively, of \citet{Maxted_ea:2015}.
Since initial metallicity and helium content are independent variables in the YaPSI tracks, this also offers the opportunity to assess the impact of $Y_0$ on the predicted mass and age.
In the runs discussed here, a ``burn-in" phase of $5000$ steps and an overall MCMC chain length of $500\,000$ steps were used. 

For the DEBs, the results obtained with the YaPSI input files are in good agreement with those of \citet{Maxted_ea:2015}: we confirm the tendency to underpredict the mass (i.e., $M_{\rm MCMC}-M_{\rm obs}<0$) in short orbital period systems and at $M_* \gtrsim 1.1\, M_\odot$. 
It should be noted that underpredicting the mass means that theoretical radii are overestimated in comparison with the observed ones, i.e., the opposite of the radius inflation typically seen in low-mass stars; these two phenomena are probably unrelated.

\begin{table*}
\begin{center}
\caption{Best-fit parameters from \texttt{BAGEMASS} for some well-studied planet-host stars (cf. table 4 of \citealt{Maxted_ea:2015}).}
\begin{tabular}{ccccccccc}
\hline
\hline
Star & $Y_0$ & $\tau_b$ [Gyr]  & $M_b$ [$M_\odot$]  & [Fe/H]$_{i,b}$  & $\chi^2$ & $\langle \tau \rangle$ [Gyr] & $\langle M \rangle$ [$M_\odot$] & $X_c$  \\
\hline
\multirow{3}{*}{HAT-P-13}  
 &  0.25   &   4.36   &   1.466   &   0.510    &    0.03  &    5.8   $\pm$  1.7   &   1.38   $\pm$   0.09   &  0.055  \\
 &  0.28   &   4.87   &   1.369   &   0.541    &    0.15  &    5.3   $\pm$  1.2   &   1.33   $\pm$   0.08   &  0.015  \\
 &  0.31   &   7.04   &   1.163   &   0.474    &    0.29  &    7.3   $\pm$  1.1   &   1.14   $\pm$   0.04   &  0.000  \\
\hline           
\multirow{3}{*}{HD209458}
 &  0.25   &   1.87   &   1.212   &   0.076    &    0.05  &    2.0   $\pm$  0.7   &   1.21   $\pm$   0.04   &  0.499  \\
 &  0.28   &   2.52   &   1.140   &   0.096    &    0.01  &    2.5   $\pm$  0.7   &   1.14   $\pm$   0.03   &  0.397  \\
 &  0.31   &   3.16   &   1.068   &   0.107    &    0.16  &    3.1   $\pm$  0.7   &   1.07   $\pm$   0.03   &  0.298  \\
\hline
\multirow{3}{*}{HD209458$^\dagger$}
 &  0.25   &   3.37   &   1.214   &   0.098    &    0.02  &    3.6   $\pm$  1.4   &   1.21   $\pm$   0.05   &  0.300  \\
 &  0.28   &   3.99   &   1.142   &   0.110    &    0.02  &    4.0   $\pm$  1.3   &   1.14   $\pm$   0.05   &  0.171  \\
 &  0.31   &   4.26   &   1.072   &   0.117    &    0.09  &    4.4   $\pm$  1.3   &   1.07   $\pm$   0.05   &  0.123  \\
\hline
\multirow{3}{*}{WASP-32}
 &  0.25   &   2.61   &   1.129   &  -0.008    &    0.02  &    2.6   $\pm$  1.1   &   1.13   $\pm$   0.04   &  0.463  \\
 &  0.28   &   3.45   &   1.054   &  -0.002    &    0.03  &    3.3   $\pm$  1.1   &   1.06   $\pm$   0.04   &  0.356  \\
 &  0.31   &   4.06   &   0.990   &   0.015    &    0.05  &    4.0   $\pm$  1.1   &   1.00   $\pm$   0.04   &  0.274  \\
\hline
\multirow{3}{*}{HD189733}
 &  0.25   &   0.18   &   0.867   &  -0.024    &    0.06  &    3.4   $\pm$  2.4   &   0.85   $\pm$   0.02   &  0.727  \\
 &  0.28   &   2.15   &   0.811   &  -0.007    &    0.03  &    4.7   $\pm$  2.8   &   0.80   $\pm$   0.02   &  0.622  \\
 &  0.31   &   5.77   &   0.753   &   0.024    &    0.03  &    6.5   $\pm$  3.1   &   0.75   $\pm$   0.02   &  0.469  \\
\hline
\multirow{3}{*}{HD189733$^\dagger$} 
 &  0.25   &  12.34   &   0.808   &   0.053    &    0.03  &   11.1   $\pm$  3.3   &   0.82   $\pm$   0.03   &  0.363  \\
 &  0.28   &  15.50   &   0.754   &   0.083    &    0.03  &   13.1   $\pm$  2.7   &   0.77   $\pm$   0.03   &  0.269  \\
 &  0.31   &  17.42   &   0.708   &   0.113    &    0.12  &   14.6   $\pm$  2.1   &   0.73   $\pm$   0.02   &  0.209  \\
\hline
\multirow{3}{*}{WASP-52}
 &  0.25   &   3.62   &   0.871   &   0.060    &    0.02  &    5.6   $\pm$  3.4   &   0.85   $\pm$   0.04   &  0.597  \\
 &  0.28   &   6.38   &   0.816   &   0.088    &    0.04  &    7.3   $\pm$  3.6   &   0.81   $\pm$   0.04   &  0.475  \\
 &  0.31   &   9.69   &   0.758   &   0.113    &    0.04  &    9.4   $\pm$  3.5   &   0.76   $\pm$   0.04   &  0.351  \\
\hline
\multirow{3}{*}{Qatar 2}
 &  0.25   &  17.43   &   0.797   &   0.180    &    1.84  &   15.6   $\pm$  1.5   &   0.81   $\pm$   0.02   &  0.300  \\
 &  0.28   &  17.49   &   0.769   &   0.247    &    5.18  &   16.2   $\pm$  1.1   &   0.78   $\pm$   0.01   &  0.261  \\
 &  0.31   &  17.49   &   0.741   &   0.303    &    9.98  &   16.6   $\pm$  0.9   &   0.75   $\pm$   0.01   &  0.223  \\
\hline
\end{tabular}
\end{center}
$^\dagger$Best-fit obtained using the observed parameters from \citet{Boyajian_ea:2015} as input.
\label{bagemass}
\end{table*}

For the planet-host stars, Table \ref{bagemass} lists the best-fitting age, mass, and initial metallicity, $\tau_b$, $M_b$, and [Fe/H]$_{i,b}$, the chi-square of the fit, $\chi^2$, the age and mass averaged over the whole MCMC chain, $\langle\tau \rangle$, $\langle M \rangle$, and the central hydrogen abundance in mass, $X_c$, which is a proxy of the evolutionary status of the best-fitting model (i.e., main sequence vs. post-main sequence).
Considering the helium enrichment relation used by \citet{Maxted_ea:2015}, $Y = 0.2485 + 0.984\, Z$, the range of observed surface metallicities for the stars in Table \ref{bagemass} ($-0.07\leq$[Fe/H]$\leq +0.41$) corresponds to a range of $Y_0$ approximately between $0.25$ and $0.31$.
The \texttt{BAGEMASS} code was thus run with the YaPSI input files for $Y_0=0.25$, $0.28$, and $0.31$ for each star.

The results reported in Table \ref{bagemass} for $Y_0=0.28$ are in good agreement with table 4 of \citet{Maxted_ea:2015}. 
In general, higher values of $Y_0$ lead to lower best-fit masses and older ages.
For HAT-P-13, the most metal-rich star in the sample, the results obtained with a higher helium abundance, $Y_0=0.31$, are a more appropriate term of comparison with \citet{Maxted_ea:2015}; in particular, we confirm that the best-fit model for this star has $X_c\approx0$, i.e., it is close to the end of the main sequence. 

The best-fitting solution found for Qatar 2 is too old to be realistic, indicating that there may be something peculiar about this star.

For HD~209458 and HD~189733, the MCMC search was also performed using the input data from the independent analysis by \citet{Boyajian_ea:2015}. 
For HD~209458, the best-fitting masses and initial metallicities found with the two data sets are almost identical, while the best-fitting ages are systematically older when the \citet{Boyajian_ea:2015} data are used.
For HD~189733, on the other hand, the best-fit models are more metal-rich, significantly less massive, and have ages close or older than the current age of the Universe. 
It should be noted that \citet{Boyajian_ea:2015} were unable to reconcile the predictions of conventional stellar models with the observational parameters of HD~189733.

\begin{figure*}
\begin{center}
\includegraphics[width=0.49\textwidth]{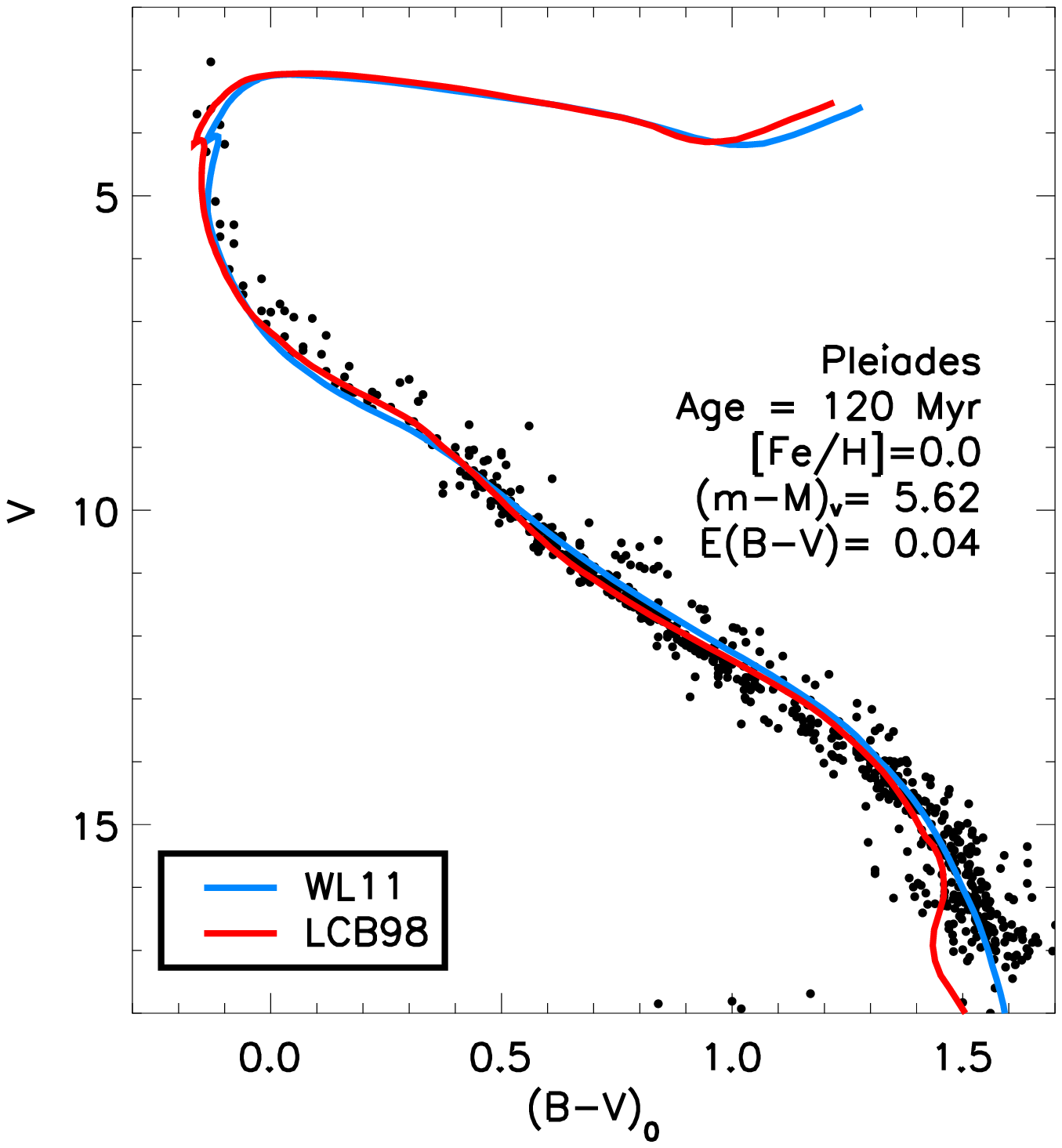}
\includegraphics[width=0.49\textwidth]{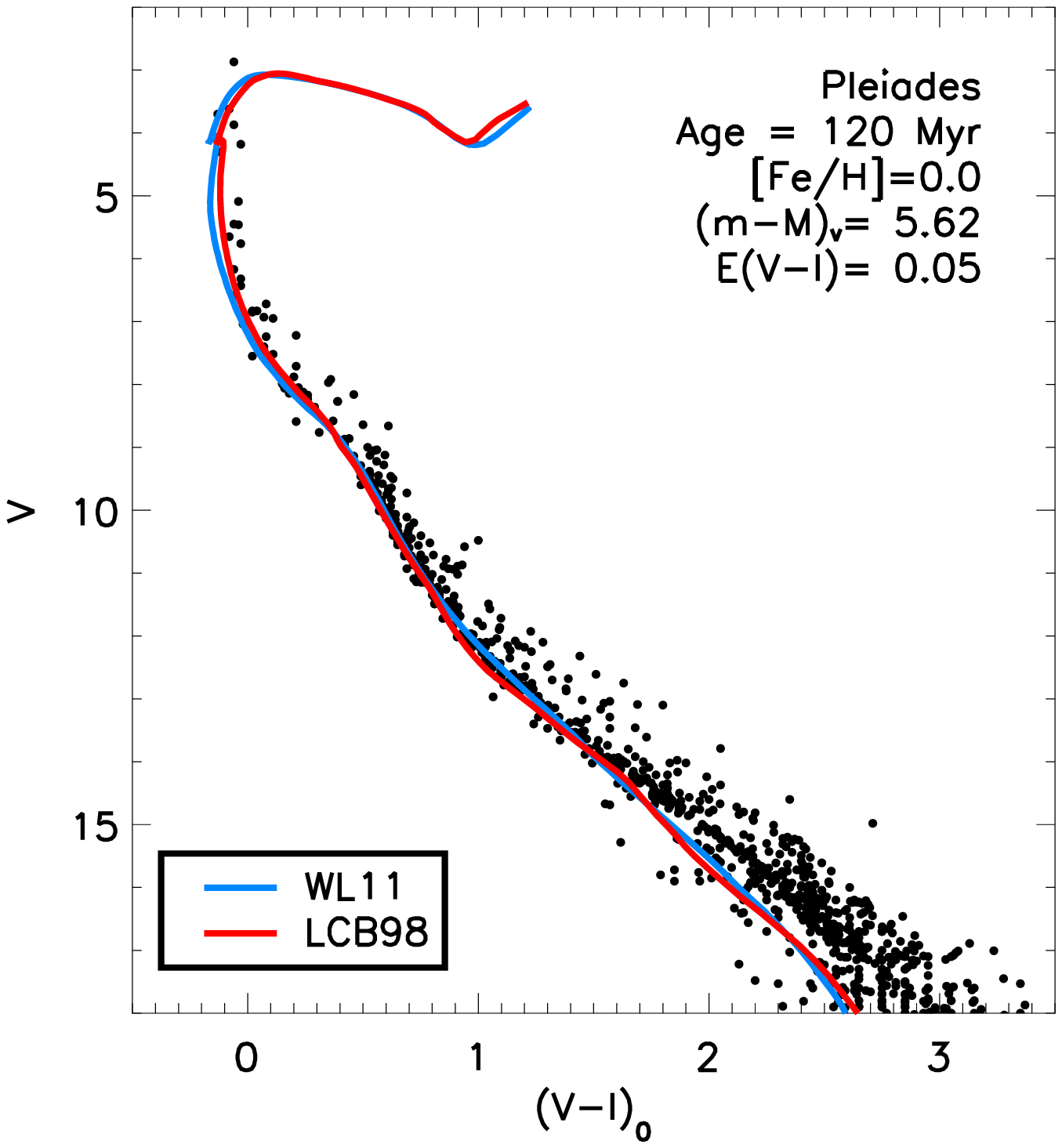}
\caption{CMDs of the Pleiades. Data from \citet{Stauffer_ea:2007,Kamai_ea:2014} (black dots); the isochrones have the following parameters: age $=120$ Myr, [Fe/H]$_0=0.0$, $Y_0=0.28$; \citet{Lejeune_ea:1997,Lejeune_ea:1998} colors (red line); \citet{Worthey_Lee:2011} colors (blue line). A good fit is obtained in the $B-V$ CMD, while both isochrones are bluer than the data for $V-I\gtrsim 1.5$ in the $V-I$ CMD. }
\label{pleiades}
\end{center}
\end{figure*}

\begin{figure*}
\begin{center}
\includegraphics[width=0.49\textwidth]{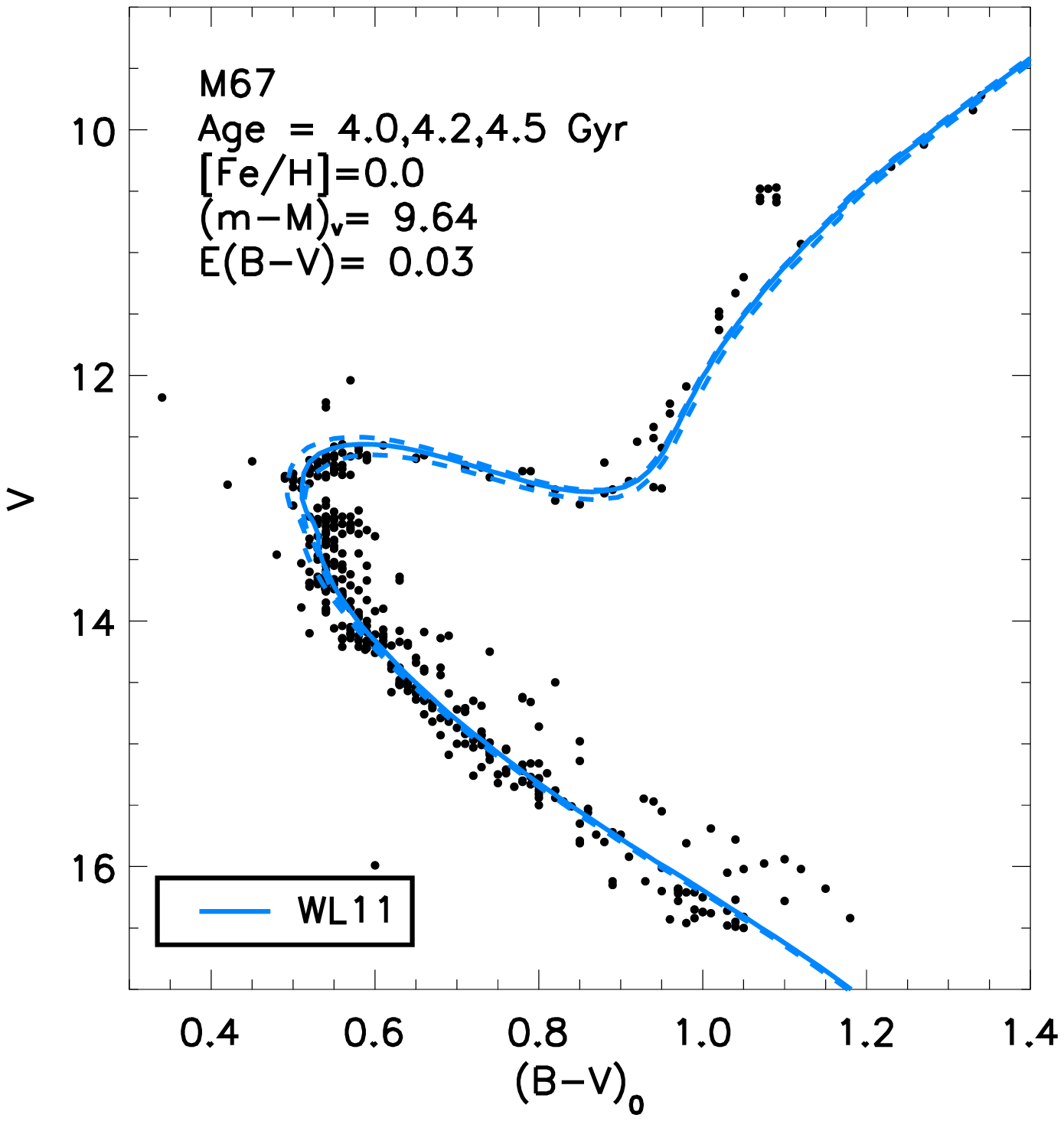}
\includegraphics[width=0.49\textwidth]{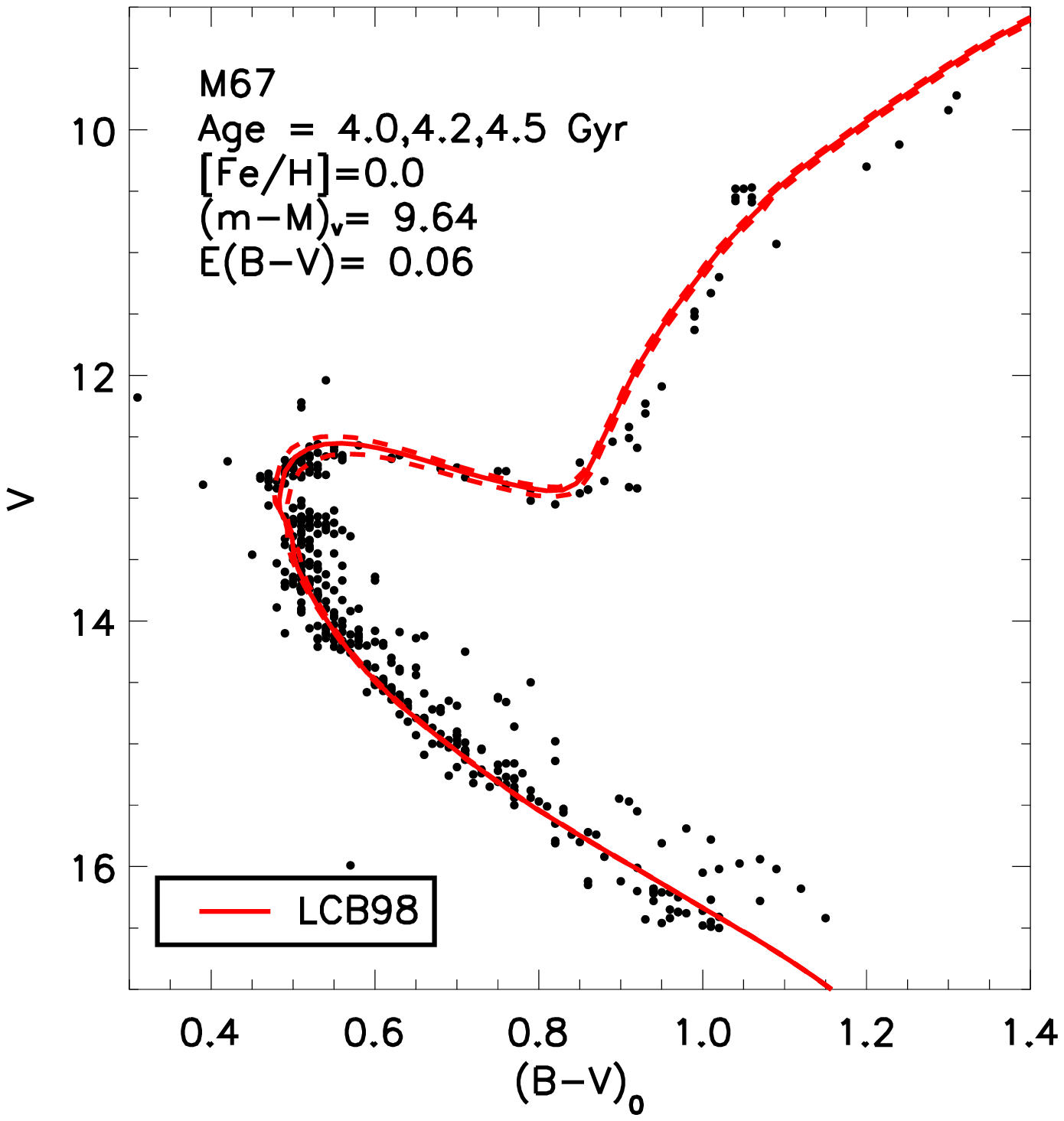}
\caption{CMDs of M67. Data from \citet{Montgomery_ea:1993,Sandquist:2004,Yadav_ea:2008, Geller_ea:2015}.
Left panel: \citet{Worthey_Lee:2011} color-calibrated isochrones at $4.2$ Gyr (solid line), and $4.0$ and $4.5$ Gyr (dashed lines), with [Fe/H]$_0= 0.0$, $Y_0=0.28$. Right panel: the same as the left panel, but using the \citet{Lejeune_ea:1997,Lejeune_ea:1998} color calibration.}
\label{m67}
\end{center}
\end{figure*}

\subsection{Isochrone fitting of stellar clusters}
\label{ssec_clusters}

Traditionally, isochrone fitting provides one of the most direct methods to test stellar evolution models, as well as to derive ages of Galactic clusters \citep[e.g.,][]{Sandage:1962b,Demarque_Larson:1964}. 
In this Section we demonstrate the capabilities of the YaPSI isochrones and the accompanying isochrone interpolation tool, by comparing them with a selection of high-quality observations of Galactic clusters.
The emphasis of this Section is therefore on testing the isochrones and the color transformations, as well as highlighting significant discrepancies which may provide fruitful avenues for further investigation and future improvement, rather than on the derivation of the best-fitting parameters of the clusters.
In the following comparisons, unless otherwise specified, the YaPSI isochrones implementing the \citet{Worthey_Lee:2011} color--$T_{\rm eff}$ calibration are used.

\subsubsection{Pleiades}

The Pleiades (M45) is often considered the quintessential population I zero-age main sequence cluster.
For this cluster we adopt the following parameters: distance modulus $(m-M)_V=5.62$ \citep[133 pc;][]{Soderblom_ea:2005,Soderblom_ea:2009,Melis_ea:2014}, reddening $E(B-V)=0.04$, $E(V-I)=0.05$ \citep{Stauffer_ea:2007}, solar metallicity \citep{Soderblom_ea:2009}, age $= 120$ Myr \citep[see][and references therein]{Dahm:2015}.
CMDs for the Pleaides constructed with data from \citet{Stauffer_ea:2007} and \citet{Kamai_ea:2014} are shown in Figure \ref{pleiades}.
In the two panels of the Figure, solar metallicity isochrones at $120$ Myr are shown, generated from the YaPSI sets with the \citet{Lejeune_ea:1997,Lejeune_ea:1998} and the \citet{Worthey_Lee:2011} color transformation, respectively.

In the $V$ vs. $B-V$ plot, the overall agreement between the isochrones and the data is satisfactory. 
Both isochrones display a moderate mismatch with the empirical single-stars main sequence for $B-V$ between $0.9$ and $1.3$.  
The \citet{Worthey_Lee:2011} color calibration is more successful at reproducing the very-low-mass end of the sequence ($B-V > 1.4$).

In the $V$ vs. $V-I$ CMD, both isochrones are bluer than the lower main sequence stars for $V-I\gtrsim 1.5$. 
This discrepancy between the models and the observations for the K and M stars in the Pleiades has been known for a long time and was noted by, e.g., \citet{Stauffer_ea:2007,Kamai_ea:2014, Choi_ea:2016}; see also \citet{Baraffe_ea:1998,Baraffe_ea:2015} for a theoretical discussion of the issue.

\subsubsection{M67}

M67 is a population I, mid-main sequence open cluster, slightly younger than the Sun (age $\approx 4$ Gyr: \citealt{Demarque_ea:1992,VandenBerg_Stetson:2004}; \citealt{Sarajedini_ea:2009} and references therein). 
An age of $4.2$ Gyr has also been derived independently using gyrochronology by \citealt{Barnes_ea:2016}.

The CMD in Figure \ref{m67} is based on the photometric data from \citet{Montgomery_ea:1993,Sandquist:2004,Yadav_ea:2008}, and the single members selection from radial velocity measurements by \citet{Geller_ea:2015}.
We adopt the distance modulus $(m-M)_V=9.74$ \citep{Sarajedini_ea:2004}, and ignore deviations from solar metallicity; in both panels, the reddening has been slightly adjusted with respect to the value of $E(B-V)=0.04$, recommended by \citet{Taylor:2007}, in order to fit the main sequence ridge.
The $\gtrsim 4.0$ Gyr isochrones with \citet{Worthey_Lee:2011} colors are also able to reproduce the shape of the main sequence turn-off and the red giant branch (left panel).
The good fit obtained for the turn-off is encouraging, since its shape is very sensitive to the convective core size, and thus to the core overshoot treatment.
The isochrones with the \citet{Lejeune_ea:1997,Lejeune_ea:1998} colors, on the other hand, are in reasonable agreement with the turn-off, but display a systematic blueward shift of the location of the red giant branch.

\subsubsection{NGC 6791}

\begin{figure}
\begin{center}
\includegraphics[width=0.49\textwidth]{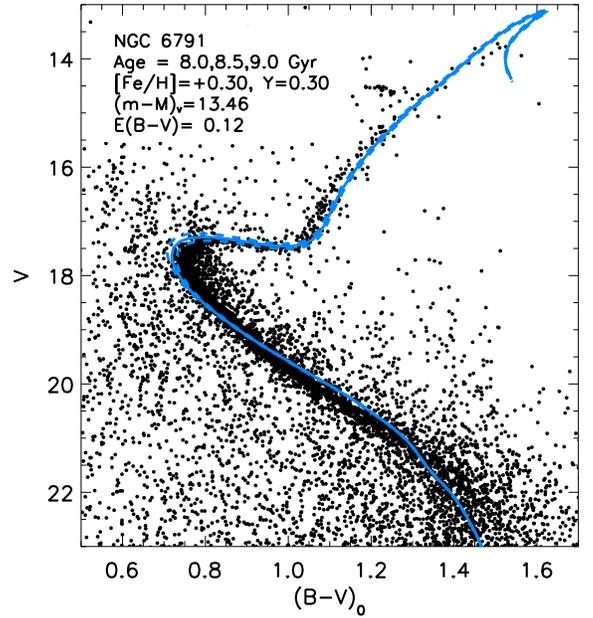}
\caption{CMD of NGC6791. Data from \citet{Brogaard_ea:2012}, implementing a differential reddening correction. The isochrones shown have [Fe/H]$_0=+0.30$, $Y_0=0.30$, ages of $8.5$ Gyr (solid line), $8.5$, and $9.5$ Gyr (dashed lines); \citet{Worthey_Lee:2011} colors.}
\label{ngc6791}
\end{center}
\end{figure}

Originally classified as a globular cluster, NGC 6791 is now considered an old, metal-rich open cluster \citep{Kinman:1965}, providing information on the old bulge population of the Milky Way.
In Figure \ref{ngc6791} we plot the photometric data of \citet{Brogaard_ea:2012}, who also performed a correction for differential reddening on the data of \citet{Stetson_ea:2003} according to the method described by \citet{Milone_ea:2012}.
From the study of an eclipsing binary member of this cluster, \citet{Grundahl_ea:2008} have estimated a distance modulus of $(m-M)_V=13.46\pm0.10$, and an age between approximately $8$ and $9$ Gyr, depending on the set of isochrones used in the comparison; this is also consistent with the more recent results of \citet{Choi_ea:2016}.

In Figure \ref{ngc6791} we plot YaPSI isochrones with [Fe/H]$_0=+0.30$ \citep[see also][]{Boesgaard_ea:2015}; $Y_0=0.30$; age $=8.0$, $8.5$, and $9.0$ Gyr.
We adopt $E(B-V)=0.12$, $(m-M)_V=13.46$. 
Although the overall fit is good, the slope of the main sequence is not reproduced uniformly well.

\subsubsection{Palomar 12}

\begin{figure}
\begin{center}
\includegraphics[width=0.49\textwidth]{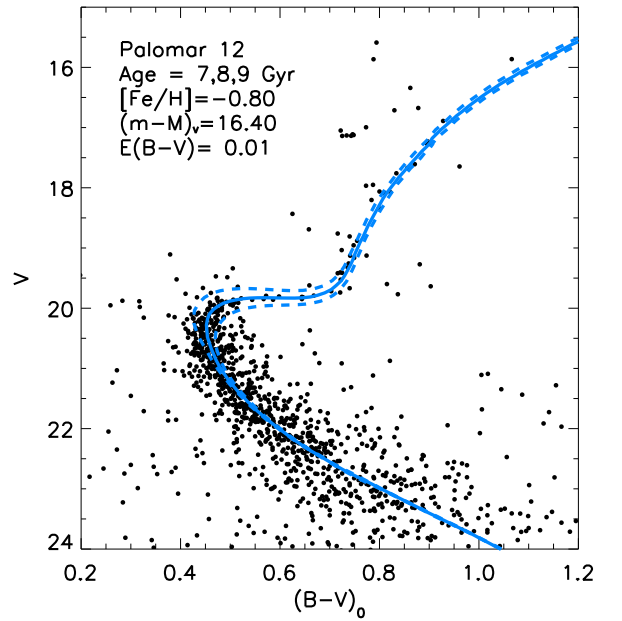}
\caption{MD of Pal 12. Data from \citet{Stetson_ea:1989}. Isochrones at $8$ Gyr (solid line) and $7$, $9$ Gyr (dashed), for [Fe/H]$=-0.8$, $Y=0.25$, \citet{Worthey_Lee:2011} color calibration.}
\label{Pal12}
\end{center}
\end{figure}

The majority of the globular clusters in the Galactic halo are very metal poor, and have non-negligible $\alpha$-element enhancement. 
As a consequence, they are outside the scope of applicability of the present YaPSI models. 
Here we focus on the younger, relatively more metal-rich cluster Palomar 12.
The composition of Palomar 12 is believed to be close to solar-scaled (i.e., [$\alpha$/Fe$]\approx 0$, see \citealt{Brown_ea:1997}), thus allowing a meaningful test of the present YaPSI isochrones.
The CMD in Figure \ref{Pal12} was constructed using data from \citet{Stetson_ea:1989}, and assuming $E(B-V)=0.02$, $(m-M)_V=16.5$, [Fe/H]$=-0.8$ (\citealt{Brown_ea:1997}; see also \citealt{Geisler_ea:2007}). 
With this choice of the parameters, our best-fit is obtained for an $8$ Gyr isochrone.
This age is younger than the $9$--$9.5$ Gyr derived by \citet{Dotter_ea:2010,Vandenberg_ea:2013}.

\section{Summary}
\label{sec_summary}

We have presented the Yale-Potsdam Stellar Isochrones (YaPSI), a new grid of stellar evolutionary tracks and isochrones covering the mass range $0.15$ to $5.00 \, M_\odot$, solar-scaled metallicity in the range [Fe/H]$_0=-1.5$ to $+0.3$, and initial helium abundance $Y_0 = 0.25$ to $0.37$.
Metallicity and helium content are assigned independently of each other, without assuming a fixed enrichment relation $\Delta Y/\Delta Z$.

The YaPSI grid is designed to provide a denser mass coverage than usually available, for the precise representation of the mass-luminosity and mass-radius relations, which is critical in the characterization of exoplanet-host stars.
 The YaPSI evolutionary tracks contain basic data, such as moments of inertia, radius at the interface between the radiative interior and the outer convection zone, and the convective turnover timescale, useful in studies of the rotational evolution of late-type stars.
For the analysis of individual stars, we provide an interface between the YaPSI tracks and an open-source \texttt{Fortran} code that performs a Monte Carlo Markov-Chain best-fit search of the observed stellar parameters. 
The YaPSI isochrones can also be applied to the classical isochrone fitting of the CMDs of open and globular clusters. 

This work is a follow-up and an extension of an initial release, described in \citet{Spada_ea:2013}.
One of the goals of both releases is to provide improved models in the low-mass range ($M\leq 0.6\, M_\odot$), to complement the widely used Yonsei-Yale isochrones \citep{Yi_ea:2001,Kim_ea:2002,Yi_ea:2003,Demarque_ea:2004}. 
For example, the models discussed here are in satisfactory agreement with the recently derived M dwarf empirical mass-luminosity relations of \citet{Benedict_ea:2016} (see Figure \ref{mlr}).

The main improvements with respect to the previous paper \citep{Spada_ea:2013} are as follows:
\begin{enumerate}
\item 
A broader range of helium content is considered; no fixed relation between helium content and metallicity content, $\Delta Y/ \Delta Z$, is assumed.
\item
The tracks are evolved from the early pre-main sequence, through the main sequence and post-main sequence up to the tip of the red giant branch.
\item
Convective core overshoot is taken into account, implementing the mass dependence of the parameter $\alpha_{\rm OV}$ derived from the semi-empirical study of \citet{Claret_Torres:2016}. 
\item 
An interface is provided between our tracks and the Bayesian stellar parameters fitting tool \texttt{BAGEMASS} \citep{Maxted_ea:2015}.
\item
The isochrone construction method has been improved; an isochrone interpolation tool in age, metallicity, and helium content is provided. 
\item
The conversion from theoretical to observational parameters (i.e., $T_{\rm eff}$--$\log g$ to magnitudes and colors) implemented in the isochrones is performed using both the \citet{Lejeune_ea:1997,Lejeune_ea:1998} and the \citet{Worthey_Lee:2011} semi-empirical calibrations. 
\end{enumerate}

\section*{Acknowledgements}
F.S. acknowledges support from the Leibniz Institute for Astrophysics Potsdam (AIP) through the Karl Schwarzschild Postdoctoral Fellowship.
YCK acknowledges support from the ``Yonsei-KASI Joint Research for the Frontiers of Astronomy and Space Science program" by the Korea Astronomy and Space Science Institute.
We thank an anonymous referee for constructive comments that led to significant improvements of the paper.
This research made use of the NASA Astrophysics Data System Bibliographic Services (ADS), and of the pspline package, provided by the National Transport Code Collaboration (NTCC). 

\appendix %

\section{The transition between low-mass and standard tracks}
\label{overlap}

In the mass range $0.60 \leq M/M_\odot \leq 1.10$, evolutionary tracks constructed with both the low-mass and the standard input physics configurations are available in the YaPSI grid (see Section \ref{ssec_massdep} for details).
In general, we find that the transition between the two subsets is fairly smooth around $M_* \gtrsim 0.9 \, M_\odot$. 
The optimal choice of this mass threshold is moderately sensitive to the chemical composition of the tracks. 
Interestingly, the strongest deviations from this general pattern are found for the chemical compositions having the most extreme values of the enrichment parameter $\Delta Y/\Delta Z$. 
These considerations are important for applications requiring high accuracy, and have been taken into account in the isochrone construction procedure described in Section~\ref{sec_isochrones}.

\begin{figure*}
\begin{center}
\includegraphics[width=0.95\textwidth]{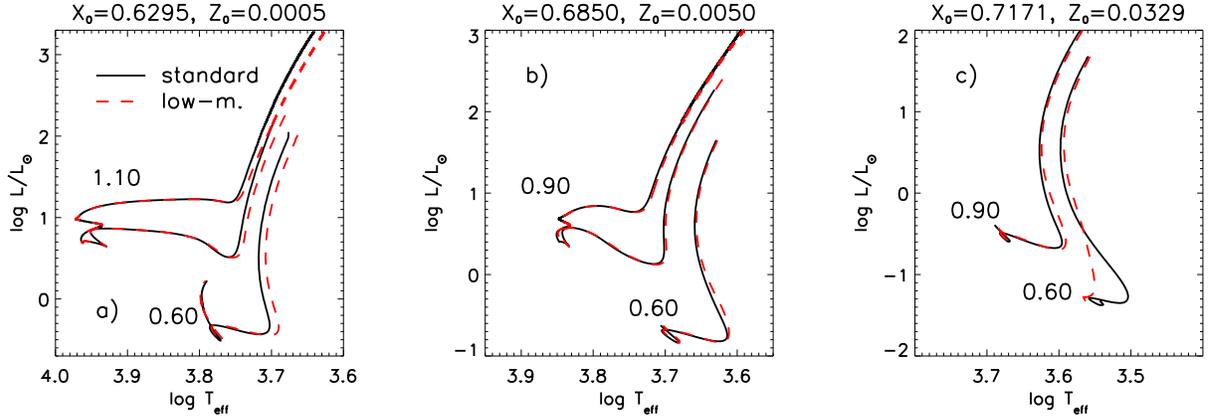}
\caption{Selected tracks illustrating the impact of the ``standard" (OPAL 2005 EOS and Eddington gray atmospheres) vs. ``low-mass" input physics configurations (SCVH EOS and atmospheric boundary conditions) at different chemical compositions. Left panel: [Fe/H]$_0=-1.5$ and $Y_0=0.37$; middle: [Fe/H]$_0=-0.5$ and $Y_0=0.31$; right: [Fe/H]$_0=+0.3$ and $Y_0=0.25$. 
The shape and the slope of the tracks are most significantly affected in the vicinity of the Hayashi line (pre-main sequence contraction and red giant phases). The evolutionary tracks are truncated at $14$ Gyr for clarity.}
\label{tracks_cont}
\end{center}
\end{figure*}

Detailed comparisons at selected compositions are shown in Figure \ref{tracks_cont}.
In the Figure, the middle panel (b) is representative of the typical, gradual transition, while the compositions of panels (a) and (c) display the two most peculiar cases.

At [Fe/H]$_0=-1.5$, $Y_0=0.37$, i.e., the most metal-poor and highest helium content available in the YaPSI grid, panel (a) illustrates the worst case scenario: both the early pre-main sequence and the red giant portions of the tracks have different slopes, and a significant discrepancy is still present even at $M_* = 1.1 \, M_\odot$.
This situation is unique in our grid.
In constructing the relative isochrones, the transition from the low-mass and the standard subgrids has been exceptionally set at $0.60 \, M_\odot$, to preserve at least the internal consistency of the intermediate age and old isochrones. 
 
Panel (b), corresponding to [Fe/H]$_0=-0.5$, $Y_0=0.31$, shows the typically modest differences at $0.60 \, M_\odot$, mostly visible along the Hayashi portion of the track and on the red giant branch; the differences are already negligibly small at $0.90 \, M_\odot$.

For [Fe/H]$_0=+0.3$, $Y_0=0.25$ (panel c), the shape of the transition from the Hayashi to the Henyey track in the pre-main sequence is qualitatively different for $M_* \lesssim 0.8 \, M_\odot$.
This case, however, is more similar to the typical behavior, in that the differences decrease with increasing mass.
This situation is thus more benign, since it can be overcome by simply choosing a larger transition mass ($\approx 1.0 \, M_\odot$).

\begin{figure*}
\begin{center}
\includegraphics[width=0.95\textwidth]{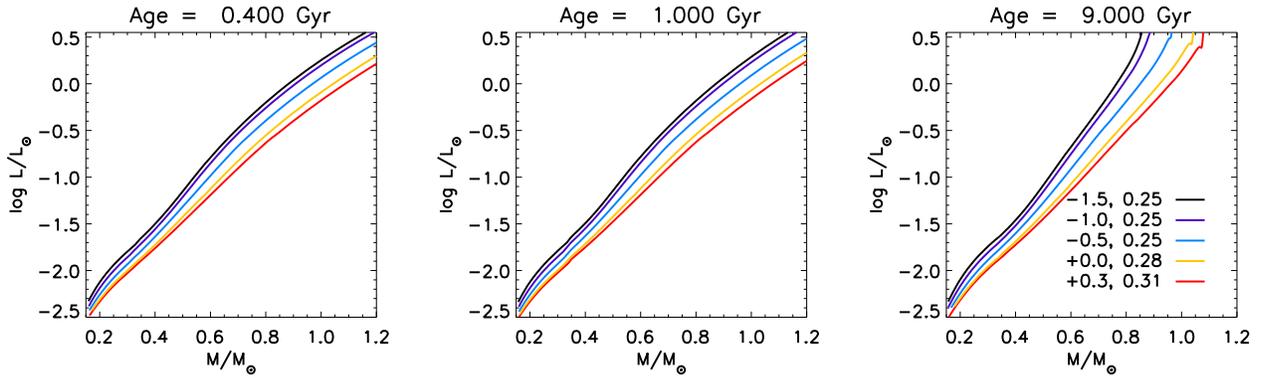}
\caption{MLRs for various compositions and ages (in each panel, the curves from top to bottom are for [Fe/H]$_0=-1.5$, $Y_0=0.25$; [Fe/H]$_0=-1.0$, $Y_0=0.25$; [Fe/H]$_0=-0.5$, $Y_0=0.25$; [Fe/H]$_0=0.0$, $Y_0=0.28$; [Fe/H]$_0=0.3$, $Y_0=0.31$; cf. legend in the plot). At $9$ Gyr (right panel), the feature at the high-mass end of most metal-rich MLRs reflects the onset of the post-main sequence phases for $M_* \gtrsim 0.9 \, M_\odot$.}
\label{mlr_cont}
\end{center}
\end{figure*}

We show that the MLRs obtained from our isochrones are satisfactorily smooth in the mass range $0.15 \leq M/M_\odot \leq 1.20$ in Figure \ref{mlr_cont}.
The most prominent deviation from smoothness is the feature near $\approx 0.35 \, M_\odot$, which is due to the transition from fully convective to solar-like interior structures (also visible in Figure \ref{mlr}).

\section{The isochrone interpolation tool}
\label{interp_test}

As part of the isochrones release, we provide a \texttt{Fortran} code\footnote{\texttt{http://www.astro.yale.edu/yapsi/interpolation\textunderscore tools.html}} that interpolates the isochrone database to user-desired age, [Fe/H]$_0$, and $Y_0$.
This code is based on the analogous tool that accompanied the YY isochrones \citep{Yi_ea:2001}. 
Interpolation between isochrones makes use of the EEPs and of the arc length $\Delta s$ concepts discussed in the previous section.

Interpolation in [Fe/H] uses a cubic polynomial; $Y_0$ interpolation is quadratic; age interpolation is linear. 
Note that the YaPSI isochrone interpolation code uses [Fe/H], instead of Z, as the interpolation variable; this choice is preferable from the numerical point of view.

\begin{figure*}
\begin{center}
\includegraphics[width=0.95\textwidth]{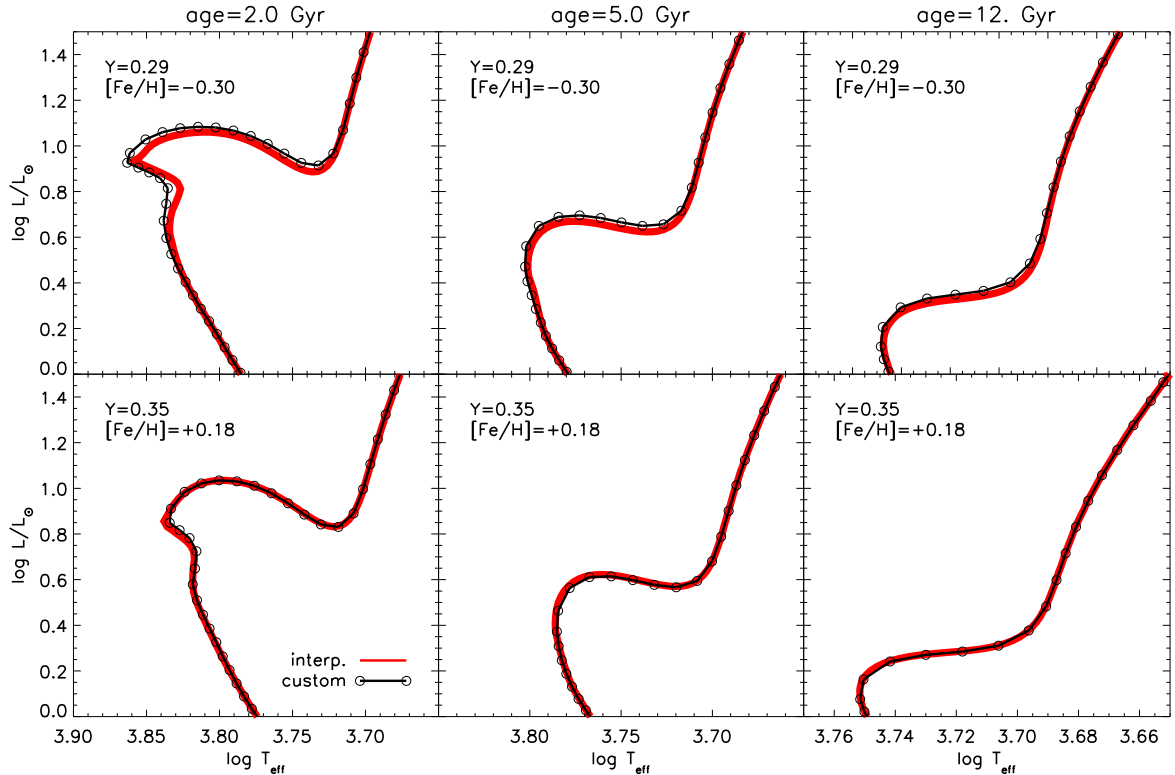}
\caption{Results of the test of the isochrone interpolation tool. Interpolated vs. custom-generated isochrones (in red and black solid lines, respectively) are shown for [Fe/H]$=-0.30$, $Y=0.29$ (upper row), and [Fe/H]$=+0.18$, $Y=0.35$ lower row.}
\label{isotest}
\end{center}
\end{figure*}

The YaPSI interpolation code is closely patterned after the analogous YY interpolation tool, which has been extensively tested since its release.
Nevertheless, to assess the accuracy of the interpolation, we have tested the code in some selected cases, by comparing its output with custom-generated isochrones, created specifically for this purpose.
The custom isochrones were constructed at $2.0$, $5.0$, and $12.0$ Gyr, from two additional sets of tracks with composition parameter [Fe/H]$=-0.30$, $Y_0=0.29$, and [Fe/H]$=+0.18$, $Y_0=0.35$, respectively, covering the mass range $0.60$--$2.20 \, M_\odot$, which were processed using the standard YaPSI isochrone construction code. 
Isochrones generated for the same parameters using the interpolation tool are compared with the custom-made ones in Figure~\ref{isotest}. 
The agreement is, in general, quite good. 
We note, however, that the largest differences between the interpolated and custom-made isochrones are found in the shape of the main sequence turn-off.
This should be taken into account for applications requiring very high accuracy.

\begin{thebibliography}{}

\bibitem[Adelberger et al.(2011)]{Adelberger_ea:2011} Adelberger, E.~G., Garc{\'{\i}}a, A., Robertson, R.~G.~H., et al.\ 2011, Reviews of Modern Physics, 83, 195 

\bibitem[Aigrain et al.(2007)]{Aigrain_ea:2007} Aigrain, S., Hodgkin, S., Irwin, J., et al.\ 2007, \mnras, 375, 29 

\bibitem[Allard et al.(1997)]{Allard_ea:1997} Allard, F., Hauschildt, P.~H., Alexander, D.~R., \& Starrfield, S.\ 1997, \araa, 35, 137 

\bibitem[Allard et al.(2011)]{Allard_ea:2011} Allard, F., Homeier, D., \& Freytag, B.\ 2011, 16th Cambridge Workshop on Cool Stars, Stellar Systems, and the Sun, 448, 91 

\bibitem[Andersen(1991)]{Andersen:1991} Andersen, J.\ 1991, \aapr, 3, 91 

\bibitem[Bahcall \& Loeb(1990)]{Bahcall_Loeb:1990} Bahcall, J.~N., \& Loeb, A.\ 1990, \apj, 360, 267 

\bibitem[Bahcall \& Pinsonneault(1992)]{Bahcall_Pinsonneault:1992} Bahcall, J.~N., \& Pinsonneault, M.~H.\ 1992, \apjl, 395, L119 

\bibitem[Baraffe et al.(1998)]{Baraffe_ea:1998} Baraffe, I., Chabrier, G., Allard, F., \& Hauschildt, P.~H.\ 1998, \aap, 337, 403 

\bibitem[Baraffe et al.(2015)]{Baraffe_ea:2015} Baraffe, I., Homeier, D., Allard, F., \& Chabrier, G.\ 2015, \aap, 577, A42 

\bibitem[Barnes(2003)]{Barnes:2003} Barnes, S.~A.\ 2003, \apj, 586, 464 

\bibitem[Barnes(2007)]{Barnes:2007} Barnes, S.~A.\ 2007, \apj, 669, 1167 

\bibitem[Barnes \& Kim(2010)]{Barnes_Kim:2010} Barnes, S.~A., \& Kim, Y.-C.\ 2010, \apj, 721, 675-685 

\bibitem[Barnes(2010)]{Barnes:2010} Barnes, S.~A.\ 2010, \apj, 722, 222  (Erratum 2001, ApJ, 729, 150)

\bibitem[Barnes et al.(2016)]{Barnes_ea:2016} Barnes, S.~A., Weingrill, J., Fritzewski, D., Strassmeier, K.~G., \& Platais, I.\ 2016, \apj, 823, 16 

\bibitem[Basu \& Antia(2008)]{Basu_Antia:2008} Basu, S., \& Antia, H.~M.\ 2008, \physrep, 457, 217 

\bibitem[Basu et al.(2010)]{Basu_ea:2010} Basu, S., Chaplin, W.~J., \& Elsworth, Y.\ 2010, \apj, 710, 1596 

\bibitem[Benedict et al.(2016)]{Benedict_ea:2016} Benedict, G.~F., Henry, T.~J., Franz, O.~G., et al.\ 2016, \aj, 152, 141 

\bibitem[Bergbusch(1993)]{Bergbusch:1993} Bergbusch, P.~A.\ 1993, \aj, 106, 1024 

\bibitem[Bessell \& Brett(1988)]{Bessell_Brett:1988} Bessell, M.~S., \& Brett, J.~M.\ 1988, \pasp, 100, 1134 

\bibitem[B{\"o}hm-Vitense(1958)]{BV58} B{\"o}hm-Vitense, E.\ 1958, Zs. Ap., 46, 108 

\bibitem[Boesgaard et al.(2015)]{Boesgaard_ea:2015} Boesgaard, A.~M., Lum, M.~G., \& Deliyannis, C.~P.\ 2015, \apj, 799, 202 

\bibitem[Bonaca et al.(2012)]{Bonaca_ea:2012} Bonaca, A., Tanner, J.~D., Basu, S., et al.\ 2012, \apjl, 755, L12 


\bibitem[Boyajian et al.(2012)]{Boyajian_ea:2012} Boyajian, T.~S., von Braun, K., van Belle, G., et al.\ 2012, \apj, 757, 112 

\bibitem[Boyajian et al.(2015)]{Boyajian_ea:2015} Boyajian, T., von Braun, K., Feiden, G.~A., et al.\ 2015, \mnras, 447, 846 

\bibitem[Bressan et al.(2012)]{Bressan_ea:2012} Bressan, A., Marigo, P., Girardi, L., et al.\ 2012, \mnras, 427, 127 

\bibitem[Brogaard et al.(2012)]{Brogaard_ea:2012} Brogaard, K., VandenBerg, D.~A., Bruntt, H., et al.\ 2012, \aap, 543, A106 

\bibitem[Brown et al.(1997)]{Brown_ea:1997} Brown, J.~A., Wallerstein, G., \& Zucker, D.\ 1997, \aj, 114, 180 

\bibitem[Burrows et al.(1989)]{Burrows_ea:1989} Burrows, A., Hubbard, W.~B., \& Lunine, J.~I.\ 1989, \apj, 345, 939 

\bibitem[Chaplin \& Miglio(2013)]{Chaplin_Miglio:2013} Chaplin, W.~J., \& Miglio, A.\ 2013, \araa, 51, 353 

\bibitem[Claret \& Torres(2016)]{Claret_Torres:2016} Claret, A., \& Torres, G.\ 2016, \aap, 592, A15 

\bibitem[Canuto(1970)]{Canuto:1970} Canuto, V.\ 1970, \apj, 159, 641 

\bibitem[Carretta \& Gratton(1997)]{Carretta_Gratton:1997} Carretta, E., \& Gratton, R.~G.\ 1997, \aaps, 121,  

\bibitem[Chabrier \& Baraffe(1997)]{Chabrier_Baraffe:1997} Chabrier, G., \& Baraffe, I.\ 1997, \aap, 327, 1039 

\bibitem[Chabrier \& Baraffe(2000)]{Chabrier_Baraffe:2000} Chabrier, G., \& Baraffe, I.\ 2000, \araa, 38, 337 

\bibitem[Charbonneau et al.(2000)]{Charbonneau_ea:2000} Charbonneau, D., Brown, T.~M., Latham, D.~W., \& Mayor, M.\ 2000, \apjl, 529, L45 

\bibitem[Chen et al.(2014)]{Chen_ea:2014} Chen, Y., Girardi, L., Bressan, A., et al.\ 2014, \mnras, 444, 2525 

\bibitem[Choi et al.(2016)]{Choi_ea:2016} Choi, J., Dotter, A., Conroy, C., et al.\ 2016, \apj, 823, 102 

\bibitem[Dahm(2015)]{Dahm:2015} Dahm, S.~E.\ 2015, \apj, 813, 108 

\bibitem[Demarque \& Larson(1964)]{Demarque_Larson:1964} Demarque, P.~R., \& Larson, R.~B.\ 1964, \apj, 140, 544 

\bibitem[Demarque \& McClure(1977)]{Demarque_McClure:1977} Demarque, P., \& McClure, R.~D.\ 1977, \apj, 213, 716 

\bibitem[Demarque et al.(1992)]{Demarque_ea:1992} Demarque, P., Green, E.~M., \& Guenther, D.~B.\ 1992, \aj, 103, 151 

\bibitem[Demarque et al.(2004)]{Demarque_ea:2004} Demarque, P., Woo, J.-H., Kim, Y.-C., \& Yi, S.~K.\ 2004, \apjs, 155, 667 

\bibitem[Demarque et al.(2008)]{Demarque_ea:2008} Demarque, P., Guenther, D.~B., Li, L.~H., Mazumdar, A., \& Straka, C.~W.\ 2008, \apss, 316, 31 

\bibitem[Dotter et al.(2008)]{Dotter_ea:2008} Dotter, A., Chaboyer, B., Jevremovi{\'c}, D., et al.\ 2008, \apjs, 178, 89-101 

\bibitem[Dotter et al.(2010)]{Dotter_ea:2010} Dotter, A., Sarajedini, A., Anderson, J., et al.\ 2010, \apj, 708, 698 

\bibitem[Dotter(2016)]{Dotter:2016} Dotter, A.\ 2016, \apjs, 222, 8 

\bibitem[Feiden \& Chaboyer(2012a)]{Feiden_Chaboyer:2012a} Feiden, G.~A., \& Chaboyer, B.\ 2012a, \apj, 757, 42 

\bibitem[Feiden \& Chaboyer(2012b)]{Feiden_Chaboyer:2012b} Feiden, G.~A., \& Chaboyer, B.\ 2012b, \apj, 761, 30 

\bibitem[Ferguson et al.(2005)]{Ferguson_ea:2005} Ferguson, J.~W., Alexander, D.~R., Allard, F., Barman, T., Bodnarik, J.~G., Hauschildt, P.~H., Heffner-Wong, A., \& Tamanai, A.\ 2005, \apj, 623, 585 

\bibitem[Ferraro et al.(1999)]{Ferraro_ea:1999} Ferraro, F.~R., Messineo, M., Fusi Pecci, F., et al.\ 1999, \aj, 118, 1738 

\bibitem[Forbes \& Bridges(2010)]{Forbes_Bridges:2010} Forbes, D.~A., \& Bridges, T.\ 2010, \mnras, 404, 1203 

\bibitem[Gallet \& Bouvier(2013)]{Gallet_Bouvier:2013} Gallet, F., \& Bouvier, J.\ 2013, \aap, 556, A36 

\bibitem[Geisler et al.(2007)]{Geisler_ea:2007} Geisler, D., Wallerstein, G., Smith, V.~V., \& Casetti-Dinescu, D.~I.\ 2007, \pasp, 119, 939 

\bibitem[Geller et al.(2008)]{Geller_ea:2008} Geller, A.~M., Mathieu, R.~D., Harris, H.~C., \& McClure, R.~D.\ 2008, \aj, 135, 2264 

\bibitem[Geller et al.(2015)]{Geller_ea:2015} Geller, A.~M., Latham, D.~W., \& Mathieu, R.~D.\ 2015, \aj, 150, 97 


\bibitem[Grevesse \& Sauval(1998)]{Grevesse_Sauval:1998} Grevesse, N., \& Sauval, A.~J.\ 1998, \ssr, 85, 161 

\bibitem[Grundahl et al.(2008)]{Grundahl_ea:2008} Grundahl, F., Clausen, J.~V., Hardis, S., \& Frandsen, S.\ 2008, \aap, 492, 171 

\bibitem[Gruyters et al.(2013)]{Gruyters_ea:2013} Gruyters, P., Korn, A.~J., Richard, O., et al.\ 2013, \aap, 555, A31 

\bibitem[Hauschildt et al.(1999)]{Hauschildt_ea:1999} Hauschildt, P.~H., Allard, F., \& Baron, E.\ 1999, \apj, 512, 377 

\bibitem[Heger et al.(2000)]{Heger_ea:2000} Heger, A., Langer, N., \& Woosley, S.~E.\ 2000, \apj, 528, 368 

\bibitem[Henry(2004)]{Henry:2004} Henry, T.~J.\ 2004, Spectroscopically and Spatially Resolving the Components of the Close Binary Stars, 318, 159 

\bibitem[Hoxie(1973)]{Hoxie:1973} Hoxie, D.~T.\ 1973, \aap, 26, 437 

\bibitem[Hubbard \& Lampe(1969)]{Hubbard_Lampe:1969} Hubbard, W.~B., \& Lampe, M.\ 1969, \apjs, 18, 297 

\bibitem[Iglesias \& Rogers(1996)]{Iglesias_Rogers:1996} Iglesias, C.~A., \& Rogers, F.~J.\ 1996, \apj, 464, 943 

\bibitem[Jeffery et al.(2013)]{Jeffery_ea:2013} Jeffery, E.~J., Platais, I., \& Williams, K.\ 2013, 18th European White Dwarf Workshop., 469, 211 

\bibitem[Kaluzny(1990)]{Kaluzny:1990} Kaluzny, J.\ 1990, \mnras, 243, 492 

\bibitem[Kamai et al.(2014)]{Kamai_ea:2014} Kamai, B.~L., Vrba, F.~J., Stauffer, J.~R., \& Stassun, K.~G.\ 2014, \aj, 148, 30 

\bibitem[Kim et al.(2002)]{Kim_ea:2002} Kim, Y.-C., Demarque, P., Yi, S.~K., \& Alexander, D.~R.\ 2002, \apjs, 143, 499 

\bibitem[Kinman(1965)]{Kinman:1965} Kinman, T.~D.\ 1965, \apj, 142, 655 


\bibitem[Lacy(1977)]{Lacy:1977} Lacy, C.~H.\ 1977, \apjs, 34, 479 

\bibitem[Lanzafame \& Spada(2015)]{Lanzafame_Spada:2015} Lanzafame, A.~C., \& Spada, F.\ 2015, \aap, 584, A30 

\bibitem[Lanzafame et al.(2016)]{Lanzafame_ea:2016} Lanzafame, A.~C., Spada, F., \& Distefano, E.\ 2016, arXiv:1609.01452 

\bibitem[Laughlin et al.(1997)]{Laughlin_ea:1997} Laughlin, G., Bodenheimer, P., \& Adams, F.~C.\ 1997, \apj, 482, 420 

\bibitem[Lee et al.(2005)]{Lee_ea:2005} Lee, Y.-W., Joo, S.-J., Han, S.-I., et al.\ 2005, \apjl, 621, L57 

\bibitem[Lejeune et al.(1997)]{Lejeune_ea:1997} Lejeune, T., Cuisinier, F., \& Buser, R.\ 1997, \aaps, 125, 229 

\bibitem[Lejeune et al.(1998)]{Lejeune_ea:1998} Lejeune, T., Cuisinier, F., \& Buser, R.\ 1998, \aaps, 130, 65 

\bibitem[Limber(1958)]{Limber:1958} Limber, D.~N.\ 1958, \apj, 127, 387 

\bibitem[Lindegren et al.(2016)]{Lindegren_ea:2016} Lindegren, L., Lammers, U., Bastian, U., et al.\ 2016, arXiv:1609.04303 

\bibitem[L{\'o}pez-Morales(2007)]{Lopez-Morales:2007} L{\'o}pez-Morales, M.\ 2007, \apj, 660, 732 

\bibitem[MacDonald \& Mullan(2013)]{MacDonald_Mullan:2013} MacDonald, J., \& Mullan, D.~J.\ 2013, \apj, 765, 126 

\bibitem[Mar{\'{\i}}n-Franch et al.(2009)]{Marin-Franch_ea:2009} Mar{\'{\i}}n-Franch, A., Herrero, A., Lenorzer, A., et al.\ 2009, \aap, 502, 559 

\bibitem[Mathieu(2000)]{Mathieu:2000} Mathieu, R.~D.\ 2000, in Pallavicini R., Micela G., Sciortino S., eds, ASP Conf. Ser. Vol. 198, Stellar Clusters and Associations: Convection, Rotation, and Dynamos. Astron. Soc. Pac., San Francisco, p. 517 

\bibitem[Maxted et al.(2015)]{Maxted_ea:2015} Maxted, P.~F.~L., Serenelli, A.~M., \& Southworth, J.\ 2015, \aap, 575, A36 

\bibitem[Melis et al.(2014)]{Melis_ea:2014} Melis, C., Reid, M.~J., Mioduszewski, A.~J., Stauffer, J.~R., \& Bower, G.~C.\ 2014, Science, 345, 1029 

\bibitem[Milone et al.(2012)]{Milone_ea:2012} Milone, A.~P., Piotto, G., Bedin, L.~R., et al.\ 2012, \aap, 540, A16 

\bibitem[Montgomery et al.(1993)]{Montgomery_ea:1993} Montgomery, K.~A., Marschall, L.~A., \& Janes, K.~A.\ 1993, \aj, 106, 181 

\bibitem[Morales et al.(2010)]{Morales_ea:2010} Morales, J.~C., Gallardo, J., Ribas, I., et al.\ 2010, \apj, 718, 502 

\bibitem[Moravveji et al.(2015)]{Moravveji_ea:2015} Moravveji, E., Aerts, C., P{\'a}pics, P.~I., Triana, S.~A., \& Vandoren, B.\ 2015, \aap, 580, A27 

\bibitem[Mowlavi et al.(2012)]{Mowlavi_ea:2012} Mowlavi, N., Eggenberger, P., Meynet, G., et al.\ 2012, \aap, 541, A41 

\bibitem[Norris(2004)]{Norris:2004} Norris, J.~E.\ 2004, \apjl, 612, L25 

\bibitem[Paczy{\'n}ski(1969)]{Paczynski:1969} Paczy{\'n}ski, B.\ 1969, Acta Astronomica, 19, 1 

\bibitem[Pence(1998)]{Pence:1998} Pence, W.\ 1998, Astronomical Data Analysis Software and Systems VII, 145, 97 

\bibitem[Penev et al.(2012)]{Penev_ea:2012} Penev, K., Jackson, B., Spada, F., \& Thom, N.\ 2012, \apj, 751, 96 

\bibitem[Perryman(2014)]{Perryman:2014} Perryman, M.\ 2014, The Exoplanet Handbook, by Michael Perryman, Cambridge, UK: Cambridge University Press, 2014,  

\bibitem[Pietrinferni et al.(2004)]{Pietrinferni_ea:2004} Pietrinferni, A., Cassisi, S., Salaris, M., \& Castelli, F.\ 2004, \apj, 612, 168 

\bibitem[Platais et al.(2003)]{Platais_ea:2003} Platais, I., Kozhurina-Platais, V., Mathieu, R.~D., Girard, T.~M., \& van Altena, W.~F.\ 2003, \aj, 126, 2922 

\bibitem[Prather(1976)]{Prather:1976} Prather, M.~J.\ 1976, Ph.D.~Thesis, Yale University

\bibitem[Press et al.(1992)]{Press:1992} Press, W.~H., Teukolsky, S.~A., Vetterling, W.~T., \& Flannery, B.~P.\ 1992 Numerical recipes in FORTRAN. The art of scientific computing (Cambridge: Cambridge University Press) 

\bibitem[Richard et al.(2002)]{Richard_ea:2002} Richard, O., Michaud, G., Richer, J., et al.\ 2002, \apj, 568, 979 

\bibitem[Rogers \& Iglesias(1995)]{Rogers_Iglesias:1995} Rogers, F.~J., \& Iglesias, C.~A.\ 1995, Astrophysical Applications of Powerful New Databases, 78, 31 

\bibitem[Rogers \& Nayfonov(2002)]{Rogers_Nayfonov:2002} Rogers, F.~J., \& Nayfonov, A.\ 2002, \apj, 576, 1064 

\bibitem[Salaris et al.(2004)]{Salaris_ea:2004} Salaris, M., Weiss, A., \& Percival, S.~M.\ 2004, \aap, 414, 163 

\bibitem[Sandage(1962a)]{Sandage:1962a} Sandage, A.\ 1962a, \apj, 135, 333 

\bibitem[Sandage(1962b)]{Sandage:1962b} Sandage, A.\ 1962b, \apj, 135, 349 

\bibitem[Sandquist(2004)]{Sandquist:2004} Sandquist, E.~L.\ 2004, \mnras, 347, 101 

\bibitem[Sarajedini et al.(1997)]{Sarajedini_ea:1997} Sarajedini, A., Chaboyer, B., \& Demarque, P.\ 1997, \pasp, 109, 1321 

\bibitem[Sarajedini et al.(1999)]{Sarajedini_ea:1999} Sarajedini, A., von Hippel, T., Kozhurina-Platais, V., \& Demarque, P.\ 1999, \aj, 118, 2894 

\bibitem[Sarajedini et al.(2004)]{Sarajedini_ea:2004} Sarajedini, A., Brandt, K., Grocholski, A.~J., \& Tiede, G.~P.\ 2004, \aj, 127, 991 

\bibitem[Sarajedini et al.(2007)]{Sarajedini_ea:2007} Sarajedini, A., Bedin, L.~R., Chaboyer, B., et al.\ 2007, \aj, 133, 1658 

\bibitem[Sarajedini et al.(2009)]{Sarajedini_ea:2009} Sarajedini, A., Dotter, A., \& Kirkpatrick, A.\ 2009, \apj, 698, 1872 

\bibitem[Saumon et al.(1995)]{Saumon_ea:1995} Saumon, D., Chabrier, G., \& van Horn, H.~M.\ 1995, \apjs, 99, 713 

\bibitem[Schwarzschild(1958)]{Schwarzschild:1958} Schwarzschild, M.\ 1958, Princeton University Press, 1958

\bibitem[Searle \& Zinn(1978)]{Searle_Zinn:1978} Searle, L., \& Zinn, R.\ 1978, \apj, 225, 357 

\bibitem[Sills et al.(2000)]{Sills_ea:2000} Sills, A., Pinsonneault, M.~H., \& Terndrup, D.~M.\ 2000, \apj, 534, 335 

\bibitem[Soderblom et al.(2005)]{Soderblom_ea:2005} Soderblom, D.~R., Nelan, E., Benedict, G.~F., et al.\ 2005, \aj, 129, 1616 

\bibitem[Soderblom et al.(2009)]{Soderblom_ea:2009} Soderblom, D.~R., Laskar, T., Valenti, J.~A., Stauffer, J.~R., \& Rebull, L.~M.\ 2009, \aj, 138, 1292 

\bibitem[Somers \& Pinsonneault(2015)]{Somers_Pinsonneault:2015} Somers, G., \& Pinsonneault, M.~H.\ 2015, \mnras, 449, 4131 

\bibitem[Southworth(2015)]{Southworth:2015} Southworth, J.\ 2015, Living Together: Planets, Host Stars and Binaries, 496, 164 

\bibitem[Spada et al.(2011)]{Spada_ea:2011} Spada, F., Lanzafame, A.~C., Lanza, A.~F., Messina, S., \& Collier Cameron, A.\ 2011, \mnras, 416, 447 

\bibitem[Spada et al.(2013)]{Spada_ea:2013} Spada, F., Demarque, P., Kim, Y.-C., \& Sills, A.\ 2013, \apj, 776, 87 

\bibitem[Stauffer et al.(2007)]{Stauffer_ea:2007} Stauffer, J.~R., Hartmann, L.~W., Fazio, G.~G., et al.\ 2007, \apjs, 172, 663 

\bibitem[Stetson et al.(1989)]{Stetson_ea:1989} Stetson, P.~B., Hesser, J.~E., Smith, G.~H., Vandenberg, D.~A., \& Bolte, M.\ 1989, \aj, 97, 1360 

\bibitem[Stetson et al.(2003)]{Stetson_ea:2003} Stetson, P.~B., Bruntt, H., \& Grundahl, F.\ 2003, \pasp, 115, 413 

\bibitem[Taylor(2007)]{Taylor:2007} Taylor, B.~J.\ 2007, \aj, 133, 370 

\bibitem[Thomas(1967)]{Thomas:1967} Thomas, H.-C.\ 1967, \zap, 67, 420 

\bibitem[Thoul et al.(1994)]{Thoul_ea:1994} Thoul, A.~A., Bahcall, J.~N., \& Loeb, A.\ 1994, \apj, 421, 828 

\bibitem[Torres et al.(2010)]{Torres_ea:2010} Torres, G., Andersen, J., \& Gim{\'e}nez, A.\ 2010, \aapr, 18, 67 

\bibitem[VandenBerg \& Stetson(2004)]{VandenBerg_Stetson:2004} VandenBerg, D.~A., \& Stetson, P.~B.\ 2004, \pasp, 116, 997 

\bibitem[VandenBerg et al.(2013)]{Vandenberg_ea:2013} VandenBerg, D.~A., Brogaard, K., Leaman, R., \& Casagrande, L.\ 2013, \apj, 775, 134 

\bibitem[VandenBerg et al.(2014)]{Vandenberg_ea:2014} VandenBerg, D.~A., Bergbusch, P.~A., Ferguson, J.~W., \& Edvardsson, B.\ 2014, \apj, 794, 72 

\bibitem[von Hippel \& Sarajedini(1998)]{vonHippel_Sarajedini:1998} von Hippel, T., \& Sarajedini, A.\ 1998, \aj, 116, 1789 

\bibitem[Worthey \& Lee(2011)]{Worthey_Lee:2011} Worthey, G., \& Lee, H.-c.\ 2011, \apjs, 193, 1 

\bibitem[Wuchterl \& Feuchtinger(1998)]{Wuchterl_Feuchtinger:1998} Wuchterl, G., \& Feuchtinger, M.~U.\ 1998, \aap, 340, 419 

\bibitem[Yadav et al.(2008)]{Yadav_ea:2008} Yadav, R.~K.~S., Bedin, L.~R., Piotto, G., et al.\ 2008, \aap, 484, 609 

\bibitem[Yi et al.(2001)]{Yi_ea:2001} Yi, S., Demarque, P., Kim, Y.-C., et al.\ 2001, \apjs, 136, 417 

\bibitem[Yi et al.(2003)]{Yi_ea:2003} Yi, S.~K., Kim, Y.-C., \& Demarque, P.\ 2003, \apjs, 144, 259 

\bibitem[Zinn(1985)]{Zinn:1985} Zinn, R.\ 1985, \apj, 293, 424 

\end{thebibliography}
\end{document}